\pdfoutput=1
\documentclass[a4paper,10pt,amsmath,amssymb,nofootinbib,onecolumn,floats,longbibliography,aps,prd]{revtex4-2}

\usepackage[british]{babel}
\usepackage{txfonts}
\usepackage{mathrsfs,bm}
\usepackage{natbib,textcase}
\usepackage{graphicx,color,soul,comment}
\usepackage{subcaption}
\usepackage{xcolor}
\usepackage[pdftex,pdfusetitle]{hyperref}
\hypersetup{colorlinks=true,linkcolor=red,citecolor=blue,filecolor=black,urlcolor=black,pdfauthor={}}
\usepackage[capitalize]{cleveref}
\usepackage{amsmath}
\usepackage{graphicx} 
\usepackage[utf8]{inputenc}
\usepackage[T1]{fontenc}
\usepackage{hhline}
\usepackage{enumitem}
\usepackage{bm}
\usepackage{mathtools,amsmath,amssymb,amsfonts,graphicx,tensor,csquotes,accents, commath,chngcntr,siunitx}
\usepackage[dvipsnames]{xcolor}
\usepackage{hyperref}
\hypersetup{colorlinks=true, citecolor=MidnightBlue,
            linkcolor=MidnightBlue, urlcolor=MidnightBlue, linktocpage=true}
\usepackage[normalem]{ulem}
\usepackage{tensor}
\usepackage{leftindex}
\usepackage{caption}
\usepackage{subcaption}
\usepackage{amssymb} 
\usepackage{orcidlink}
\usepackage{pifont}
\usepackage{lmodern}
\usepackage{cancel}

\usepackage{booktabs}
\usepackage{float}

\usepackage{array}

\usepackage{xcolor}

\usepackage{tabularray}

\DeclareMathAlphabet{\mathpzc}{OT1}{pzc}{m}{it}
\newcommand{\MP}{M_{\rm P}}

\begin{document}

\title{The price of regularity: core gradient instability in nonsingular dyonic black holes}

\author{Lodovico Capuano ~\orcidlink{0009-0001-0369-6635}}
\email{lodovico.capuano@uniroma1.it}
\affiliation{Dipartimento di Fisica, Sapienza Università di Roma, Piazzale Aldo Moro 5, 00185, Roma, Italy}

\author{Alessandro Longo ~\orcidlink{0009-0006-8557-7344}}
\email{alongo@apc.in2p3.fr}
\affiliation{Universit\'e Paris Cit\'e, CNRS, AstroParticule et Cosmologie, 10 rue Alice Domon et L\'eonie Duquet, 75013 Paris, France}

\begin{abstract}
Spacetime singularities displayed by classical black hole solutions are expected to be cured by high-energy effects, beyond the regime of validity of General Relativity. On the other hand, there are well-known examples of classical theories allowing for nonsingular black hole solutions. The most prominent is obtained by complementing the gravitational sector of General Relativity with a minimally coupled nonlinear-electrodynamics sector. It has recently been shown that purely electric/purely magnetic nonsingular black holes in an Einstein-nonlinear-electrodynamics setup are always plagued by a gradient instability. In this work, we derive the quadratic action for linear perturbations of static, spherically symmetric dyonic solutions of Einstein-nonlinear-electrodynamics
theories with a general dependence on both invariants. We show that regularity of the center,
absence of ghosts and angular gradient stability cannot hold simultaneously: arbitrarily close
to a regular center, at least one photon polarization has a negative squared angular propagation speed. We further show that regularity drives one electromagnetic kinetic eigenvalue to zero at the center, which may limit perturbative control
near the core. Evading the obstruction requires extending the theory, for example by introducing additional degrees of freedom.
\end{abstract}

\maketitle

\section{Introduction}


Classical black hole (BH) solutions in General Relativity (GR) typically exhibit curvature singularities, signaled by the divergence of curvature invariants as the singular region is approached~\cite{Schwarzschild:1916uq,Reissner:1916cle,Kerr:1963ud,Newman:1965my}. Moreover, the classical theorems of GR show that the formation of singularities in a dynamical gravitational collapse is inevitable, provided that the matter-energy content of the spacetime fulfills the classical energy conditions~\cite{Penrose:1964wq,Hawking:1966vg,Hawking:1970zqf,Hawking:1973uf,Senovilla:2014gza}. This phenomenon is interpreted as a manifestation of the effective-field theory (EFT) nature of GR itself~\cite{Donoghue:1994dn,Burgess:2003jk}. In this perspective, one generally expects such singularities to be resolved in an ultraviolet (UV)-complete theory of gravity~\cite{Wald:1980jn,Horowitz:1995gi,Hofmann:2015xga,Casals:2016odj,Gupta:2019cmo,Crowther:2021qij}. While the question of finding a consistent UV-completion of GR is still open, it is interesting to explore classical extensions thereof that may admit regular BH solutions. 


In practice, BH singularities can be resolved entirely on the gravitational side, without any additional matter content. By suitably tuning an infinite tower of higher-curvature corrections to the Einstein-Hilbert action – chosen so as to preserve second-order field equations on spherically symmetric backgrounds – one obtains asymptotically flat, everywhere-regular BH solutions in a spacetime dimension
greater than or equal to five~\cite{Bueno:2024dgm}, later generalized to Lagrangians depending on all the basic curvature invariants~\cite{Frolov:2024hhe} and shown to arise dynamically from gravitational collapse~\cite{Bueno:2024eig}. Only more recently has an analogous construction been extended to four-dimensional spacetimes~\cite{Bueno:2025zaj}.

On the other hand, it has been shown that nonsingular solutions can be found introducing a nonlinear electrodynamics (NLED) content in the theory, even without altering the GR gravitational sector~\cite{Ayon-Beato:1998hmi,Ayon-Beato:1999kuh,Ayon-Beato:2000mjt,Bronnikov:2000vy,Bronnikov:2022ofk,Ansoldi:2008jw,Dymnikova:2004zc,Balart:2014cga,Balart:2014jia,Fan:2016hvf,Rodrigues:2018bdc,AraujoFilho:2026hun}. This setup will be the focus of the present manuscript. Although different classes of regular BHs have been shown to exist in NLED, it has never been established in the most general case whether such objects are linearly stable. In~\cite{DeFelice:2024seu} it was shown that, in a subset of NLED, where the Lagrangian only depends on the parity-even electromagnetic invariant $X$, purely electric and purely magnetic regular BHs are generally affected by a gradient instability in the angular sector, localized around the would-be regular center. This construction, however, structurally excludes dyonic solutions: for a Lagrangian $\mathcal{L}(X)$, demanding regularity of the metric at the center forces the product of the electric and magnetic charge to vanish, irrespective of the weak-field behavior of $\mathcal{L}$.
Regular dyonic BHs -- exhibiting simultaneously nonvanishing electric and magnetic charge -- only become possible once the Lagrangian is allowed to depend also on the parity-odd invariant $Y$, i.e. $\mathcal{L}(X,Y)$~\cite{Bokulic:2025brf,Tsuda:2023tcs,Bronnikov:2026kgl}. 

We stress that this generalization is not primarily tied to abandoning the Maxwellian weak-field limit: explicit dyonic solutions have been constructed that recover the standard Maxwell Lagrangian at weak field~\cite{Bokulic:2025brf}, at the price, however, of an intrinsic, nontrivial relation entangling the BH mass with its electric and magnetic charges. The stability of this genuinely dyonic sector, in which the pseudoscalar invariant $Y$ does not vanish on the background, has so far remained completely unexplored for regular BHs, since all existing perturbative stability analyses~\cite{Moreno:2002gg,Nomura:2020tpc,DeFelice:2024seu} have been restricted to configurations on which $Y$ vanishes on the background. More general dyonic BHs with scalar hair have been studied in $U(1)$-gauge-invariant scalar-vector-tensor theories~\cite{Taniguchi:2025bmc}, where general no-ghost and gradient-stability conditions were derived and applied to several extended Einstein-Maxwell-scalar models. These analyses, however, did not address the compatibility between linear stability and regularity of the BH core.

In the present work, we fill this gap in the recent literature, exploring the conditions for the absence of gradient instabilities in regular dyonic BHs.

In Sec.~\ref{sec:background}, we describe the NLED setup, specializing to static solutions that exhibit no singularity at the core $r = 0$. We also introduce the scalings and characteristic exponents of relevant quantities close to the core.

In Sec.~\ref{sec:linear_pert}, we derive the quadratic action governing the dynamics of linear perturbations, introducing a suitable gauge choice and an appropriate parametrization of physical degrees of freedom.

In Sec.~\ref{sec:stability}, we work out the no-ghost conditions and study the gradient-stability close to the nonsingular core at $r = 0$. 

\emph{Notation and conventions:}

We adopt natural units $c = \hbar= 1$, and the mostly-plus signature $(-,+,+,+)$ for the metric tensor.

\section{Nonlinear electrodynamics: theoretical setup}\label{sec:background}

In this section, we provide the general theoretical setup of our analysis, i.e. NLED on a curved background. 

The first explicit realization of NLED was proposed by Born and Infeld~\cite{Born:1934gh}, who introduced a square-root Lagrangian aimed at regularizing the electron self-energy by imposing a maximal field strength, building on Born's earlier proposal of a modified electromagnetic theory with finite electron radius~\cite{Born:1933qff}. A microscopic origin for nonlinear interactions quartic in the electromagnetic field strength was later provided by Heisenberg, Euler, and Weisskopf~\cite{Heisenberg:1936nmg,Weisskopf:1936hya}, who derived the one-loop effective action of quantum electrodynamics in a constant background field, with the resulting nonlinearities encoding vacuum-polarization effects. NLED also arises as a low-energy effective description in string theory~\cite{Fradkin:1985qd}. In the context of BH physics, NLED attracted renewed interest after it was shown that suitable nonlinear electromagnetic sources can support nonsingular spacetime geometries~\cite{Ayon-Beato:1998hmi,Ayon-Beato:1999kuh,Ayon-Beato:2000mjt,Bronnikov:2000vy,Bronnikov:2022ofk,Ansoldi:2008jw,Dymnikova:2004zc,Balart:2014cga}. A common strategy in this literature is to prescribe a regular metric profile and reconstruct the NLED Lagrangian from the field equations, effectively adopting a reverse-engineering approach~\cite{Cadoni:2022chn,Cadoni:2023nrm}. In doing so, the nonlinear electromagnetic theory is typically treated as a classical model valid throughout the geometry, including the high-field regime near the core, rather than as a low-energy EFT with a specified finite cutoff. We will initially adopt this viewpoint and return in the final section to the interpretation of our results from an EFT perspective.

\subsection{Equations of motion}
The full action of NLED, in general, reads
\begin{equation}
    S=\int {\rm d}^4x\,\sqrt{-g}\,\left[\frac{\MP^2}{2}R+\mathcal{L}\left(X,Y\right)\right]\,,
    \label{eq:background_action}
\end{equation}
where $R$ is the Ricci scalar, $\MP$ the reduced Planck mass, $g$ the determinant of the metric tensor $g_{\mu\nu}$ and $\mathcal{L}(X,Y)$ a generic nonlinear function of the electromagnetic invariants $X$ and $Y$. We define the two invariants $X$ and $Y$, respectively even- and odd-parity, according to the following convention
\begin{equation}
     X\equiv -\frac{1}{4}F_{\mu \nu}F^{\mu \nu}\>,\> Y\equiv -\frac{1}{4}F_{\mu \nu}\tilde{F}^{\mu \nu}\,,
      \label{eq:EM_invariants0}
\end{equation}
where $F_{\mu\nu}$ is the field strength of the $U(1)$-gauge field $A_{\mu}$, and $\tilde F_{\mu\nu}$ is its Hodge dual
\begin{equation}
    F_{\mu\nu}=\nabla_{\mu}A_{\nu}-\nabla_{\nu}A_{\mu}\,,\qquad\qquad\tilde{F}^{\mu\nu}=\frac{1}{2}e^{\mu\nu\rho\sigma}F_{\rho\sigma},
    \label{eq:field_strength}
\end{equation}
$e^{\mu \nu\rho\sigma}$ being the Levi-Civita tensor. 

We will focus our attention on static and spherically symmetric configurations for the background metric 
\begin{equation}
    {\rm d}s^2=g_{\mu\nu}{\rm d}x^\mu {\rm d}x^\nu=-f(r){\rm d}t^2+\frac{1}{h(r)}{\rm d}r^2+r^2 ({\rm d}\theta^2+\sin^2\theta\,{\rm d} \phi^2)
    \label{eq:metric-ansatz}
\end{equation}
together with the following profile for the gauge field
\begin{equation}
    A_{\mu}{\rm d}x^\mu=A_0(r){\rm d}t-q_M \cos{\theta}\,{\rm d}\phi\,,
\end{equation}
where $q_M$ corresponds to a magnetic charge. This charge is forced to be constant ($q_M'(r)= 0$) in the bulk by the Bianchi identity $\nabla_\mu \tilde F^{\mu\nu}=0\,$, which automatically follows from Eq.~\eqref{eq:field_strength}, i.e. by requiring that the electromagnetic field strength be a closed two-form.

The two electromagnetic invariants read
\begin{equation}
    X=\frac{h}{f}\frac{A_0'^2}{2}-\frac{q_M^2}{2r^4}\,,\qquad\qquad Y=-\sqrt{\frac{h}{f}}\frac{A_0'q_M}{r^2}\,.
    \label{eq:EM_invariants}
\end{equation}
while the equations of motion, obtained by varying Eq.~\eqref{eq:background_action} with respect to $g^{\mu\nu}$ and $A^\mu$, are given by the Einstein-NLED equations
\begin{equation}
    \MP^2 G_{\mu\nu}=T_{\mu\nu}\,,\qquad\qquad \nabla_\mu \left(\mathcal{L}_XF^{\mu\nu}+\mathcal{L}_Y\tilde F^{\mu\nu}\right)=0\,,
    \label{eq:EinsteinMaxwell}
\end{equation}
where we introduced the Einstein tensor $G_{\mu\nu}=R_{\mu\nu}-g_{\mu\nu}R/2$ ($R_{\mu\nu}$ and $R$ being respectively the Ricci tensor and Ricci scalar), and we defined the stress-energy tensor
\begin{equation}
    T_{\mu\nu} =-\frac{2}{\sqrt{-g}}\frac{\delta \left(\sqrt{-g}\mathcal{L}\right)}{\delta g^{\mu\nu}}=\mathcal{L}_X F_{\mu \alpha}F_\nu^{\,\,\alpha}+\left(\mathcal{L}-Y\mathcal{L}_Y\right)g_{\mu\nu}\,.
\end{equation}
On the other hand, the covariant conservation of the stress-energy tensor $\nabla_\mu T^{\mu\nu}=0$, or equivalently the gravitational Bianchi identity $\nabla_\mu G^{\mu\nu}=0$, is automatically satisfied on shell. We also indicated with the subscripts $X$ and $Y$ the derivatives with respect to the corresponding electromagnetic invariants. With this notation, the Maxwell limit is recovered as $\mathcal{L}_X\to 1$ and $\mathcal{L}_Y\to 0$.

The $tt$ and $rr$ components of the Einstein equation read as follows
\begin{equation}
\begin{split}
    &\MP^2G_{tt}=\MP^2f\left(\frac{1-h}{r^2}-\frac{h'}{r}\right)=h\mathcal{L}_X A_0'(r)^2-f\left(\mathcal{L}-Y \mathcal{L}_Y\right)\,,\\
    &\MP^2G_{rr}=\MP^2\left(\frac{h-1}{hr^2}+\frac{f'}{fr}\right)=-\frac{\mathcal{L}_X A_0'(r)^2}{f}+\frac{\left(\mathcal{L}-Y \mathcal{L}_Y\right)}{h}\,.
\end{split}
\label{eq:Einstein_ttrr}
\end{equation}
Combining the $tt$ and $rr$ components of Eq.~\eqref{eq:Einstein_ttrr}, one finds
$h f'- h'f=0$ and therefore $f=C h$, with $C$ an integration constant. The standard normalization of the
time coordinate allows one to fix $C=1$.
Hence, in the following, we set
$f=h$.
With this condition, the third independent diagonal component of the Einstein equation becomes
\begin{equation}
    \frac{\MP^2}{r^2}G_{\theta \theta}=\frac{\MP^2}{r^2\sin^2\theta}G_{\phi \phi}=\MP^2\left(\frac{f''}{2}+\frac{f'}{r}\right)=\mathcal{L}-Y \mathcal{L}_Y+\frac{q_M^2}{r^4}\mathcal{L}_X\,.
\label{eq:Einstein_thetatheta}
\end{equation}
Finally, the nontrivial $t$ component of the gauge-field equation reads
\begin{equation}
    \frac{\partial}{\partial r}\left(r^2 \mathcal{L}_XA_0'-q_M\mathcal{L}_Y \right)=0\,,
\end{equation}
which is solved by
\begin{equation}
    A_0'(r)=\frac{q_E+q_M\mathcal{L}_Y}{\mathcal{L}_X r^2}\,,
    \label{eq:charge_cons}
\end{equation}
with the integration constant $q_E$ being identified as an electric charge. In the Maxwell limit this simply becomes the Coulomb law $A'_0(r)= q_E/r^2$. Eq.~\eqref{eq:charge_cons} can be interpreted as the electric charge conservation equation, as the Bianchi identity is identified as the magnetic charge conservation equation. 

\subsection{Nonsingular solutions and core scalings}
We now want to derive the scaling rules for relevant objects introduced in the previous subsection, close to the coordinate origin $r=0$, when regularity is required there. Regularity is imposed by requiring finiteness of the curvature invariants. As an example, one can compute the Kretschmann scalar
\begin{equation}
    R^{\mu\nu\rho\sigma}R_{\mu\nu\rho\sigma}=\frac{4(1-f)^2}{r^4}+\frac{4(f')^2}{r^2}+(f'')^2\,.
    \label{eq:Kretschmann}
\end{equation}
and notice that, since it is made of a sum of squares, regularity around $r=0$ requires the three combinations
\begin{equation}\label{eq:scalingf}
  \frac{1-f}{r^2}\>,\> \frac{f'}{r}\>,\> f''
\end{equation}
to remain finite near the core.
Notice that these relations do not rely on the existence of a Taylor expansion for $f$ at $r=0$. When a leading power-law behavior exists, we may parametrize $1-f\sim r^\alpha$, with $\alpha\ge2$. These conditions can also be realized by nonanalytic functions, including profiles that vanish faster than any finite power of $r$.\footnote{For functions that are flat at the origin, such as $e^{-1/r}$, the power-law parametrization $r^\alpha$ may be dropped altogether, since $1-f$ vanishes faster than any finite power of $r$ as $r\to0$.}
Notice that the conditions~\eqref{eq:scalingf} ensure finiteness of any scalar polynomial invariants algebraically constructed from the Riemann tensor. Indeed, static spherical symmetry implies that
the independent components of the Riemann tensor are controlled by
the three combinations
$f''$ $f'/r$ and $(1-f)/r^2$.
Therefore, boundedness of the Kretschmann scalar implies boundedness
of each independent Riemann component and hence of any scalar
polynomial constructed algebraically from the Riemann tensor
\cite{Bronnikov:2022ofk}.

Using the Einstein equations, it is possible to derive a set of scaling conditions for the electromagnetic quantities. In particular, summing Eq.~\eqref{eq:Einstein_ttrr} and~\eqref{eq:Einstein_thetatheta} yields 
\begin{equation}
    r^2 f''+2(1-f)=\frac{2 r^2 \mathcal{L}_X}{\MP^2}\left(A_0'^2+\frac{q_M^2}{r^4}\right)\,.
    \label{eq:Einstein_combined}
\end{equation}
When a leading power-law behavior exists, the LHS of Eq.~\eqref{eq:Einstein_combined} scales as $r^{\beta}\>,\>\beta \geq \alpha$ near $r= 0$. The same behavior must therefore be obeyed by the RHS

\begin{equation}
    \mathcal{L}_X\left(A_0'^2+\frac{q_M^2}{r^4}\right)=\frac{1}{r^4}\left(\frac{(q_E+q_M\mathcal{L}_Y)^2}{\mathcal{L}_X}+q_M^2\mathcal{L}_X\right)\sim r^{\beta-2}\,.
    \label{eq:LX_scaling}
\end{equation}
Notice that the sign of the LHS is dictated by the sign of $\mathcal{L}_X$. In particular, the parenthesis consists of a sum of two objects with the same sign. As a consequence, no cancellation between them is possible; therefore each must vanish at least as $r^{\nu}\>,\>\nu\geq \beta+2$. More explicitly, regularity at $r=0$ implies \begin{equation}
    \mathcal{L}_X\sim r^q\>,\> q_E+q_M \mathcal{L}_Y
\sim r^n\>,\quad q\geq \beta+2\>,\>n\geq (\beta+2+q)/2\end{equation}
Notice that, since $q\geq 4$, $\mathcal{L}_X$ is forced to vanish as $r\to0$. This means that a magnetically charged regular BH cannot remain in the Maxwell regime near the center. By contrast, $\mathcal{L}_Y\to -q_E/q_M$ near a regular dyonic core. 


\section{Linear Perturbations}
\label{sec:linear_pert}

In this section we derive the quadratic action for linear perturbations on top of a spherically symmetric background. We begin by perturbing the metric and gauge field as $g_{\mu \nu}\to g_{\mu\nu}+h_{\mu \nu}$ and $A_{\mu}\to A_{\mu}+\delta A_{\mu}$ respectively, and decompose the fluctuations in spherical harmonics. 

The technical details of this procedure -- the gauge-fixing and the subsequent manipulation of the action -- are collected in Secs.~\ref{subsec:gauge_choice} and~\ref{subsec:action_manipulation}. Since the derivation follows standard techniques~\cite{Kobayashi:2012kh,Kobayashi:2014wsa}, the reader already familiar with them may proceed directly to Sec.~\ref{subsec:matrices}, where the resulting mode-coupling matrices are presented.

\subsection{Perturbation decomposition and gauge choice}

\label{subsec:gauge_choice}
It is convenient to perform a 2+2 decomposition and distinguish the non-angular components (lower case latin indices) from the angular ones (upper case latin indices). With this convention, the nontrivial components of the background metric are $g_{ab}$ and $g_{AB}=r^2\Omega_{AB}$ ($\Omega_{AB}$ being the metric on the unit-sphere) and we can write\footnote{In the following equations, a sum over the angular-momentum labels is understood.}
\begin{equation}h_{ab}=H_{ab}^{\ell m}Y^{\ell m}\>,\>h_{aA}=w_a^{\ell m}Y_A^{\ell m}+z_a^{\ell m}X_A^{\ell m}\>,\>h_{AB}=r^2(K^{\ell m}\Omega_{AB}Y^{\ell m}+G^{\ell m}Y_{AB}^{\ell m}+h_2^{\ell m}X_{AB}^{\ell m})
\end{equation}
with \begin{equation}
    H_{ab}=\begin{bmatrix}
        fH_0&H_1\\
        H_1&H_2/f
    \end{bmatrix}
\end{equation}
and where\footnote{The scalar spherical harmonics are defined by $\int_{S^2}Y^{\ell m}(Y^{\ell'm'})^{*}=\delta_{\ell,\ell'}\delta_{m,m'}\>,\>d_A d^AY^{\ell m}=-\ell(\ell+1)Y^{\ell m}\equiv -L Y^{\ell m}$.}
\begin{equation}
    Y_A^{\ell m}=\partial_A Y^{\ell m}\equiv d_AY^{\ell m}\>,\>X_A^{\ell m}=e_{AB}d^BY^{\ell m}\>,\>Y_{AB}^{\ell m}=d_{\langle A}Y_{B\rangle}^{\ell m}\>,\>X_{AB}=d_{(A}X_{B)}^{\ell m}
\end{equation}
with $d_A$ and $e_{AB}$ the covariant derivative (associated with the Levi-Civita connection) and the Levi-Civita tensor on $S^2$. Analogously, we can write
\begin{equation}
    \delta A_a=a_a^{\ell m}Y^{\ell m}\>,\>\delta A_A=a^{\ell m}Y_A^{\ell m}+b^{\ell m}X^{\ell m}_A 
\end{equation}
Since the physical background is spherically symmetric, all $m$-modes at fixed $\ell$ are equivalent. We may therefore set $m=0$ without loss of generality, keeping in mind that the monopole gauge potential itself is rotationally invariant only up to a $U(1)$ gauge transformation. Moreover, we will henceforth omit the angular-momentum labels whenever no confusion can arise. According to their transformation properties under the parity transformation $(\theta,\phi)\to (\pi-\theta,\phi+\pi)$, the perturbations $\{H_{ab},w_a,K,G,a_a,a\}$ belong to the even-parity sector, whereas $\{z_a,h_2,b\}$ belong to the odd-parity sector. We stress, however, that this classification is purely kinematical and does not imply that the two sectors are dynamically independent. In the presence of a magnetic charge and, more generally, on a dyonic background or for parity-violating interactions, even- and odd-parity perturbations can mix already at quadratic order. Since we will be interested in this general situation, we retain both sectors throughout the analysis and do not assume that the quadratic action is block diagonal in parity. Before turning to the dynamics, we briefly characterize the gauge redundancy of the harmonic variables under infinitesimal diffeomorphisms and $U(1)$ gauge transformations. Under an infinitesimal diffeomorphism $x^\mu\to x^\mu-\xi^\mu$, the metric and gauge-field perturbations transform as $
    \delta h_{\mu\nu}=\nabla_\mu\xi_\nu+\nabla_\nu\xi_\mu\>,\>
    \delta A_\mu = A_\nu\nabla_\mu\xi^\nu
    +\xi^\nu\nabla_\nu A_\mu
    =
    \partial_\mu\left( A_\nu\xi^\nu\right)
    +\xi^\nu F_{\nu\mu},$
where $\xi_\mu=\left(
    \xi_a^\ell Y^{\ell,0},
    \,
    \xi_S^\ell Y_A^{\ell,0}+\xi_V^\ell X_A^{\ell,0}
    \right)\>,\>
    \xi_a^\ell 
    =
    \left(
    f\xi_0^\ell ,\xi_1^\ell /f
    \right)$. More explicitly
\begin{equation}
    \begin{aligned}
        &\delta H^\ell _{ab}=D_a \xi^\ell _b +D_b \xi^\ell _a\>,\>\delta w_a Y^{\ell,0}_A+\delta z_a X^{\ell,0}_A=(r^2\partial_a( \xi^\ell _S/r^2)+\xi^\ell _a)Y^{\ell,0}_A+(r^2\partial_a( \xi^\ell _V/r^2))X^{\ell,0}_A \\
        &r^2(\delta K^\ell  \Omega_{AB}Y^{\ell,0}+\delta G^\ell  Y^{\ell,0}_{AB}+\delta h^\ell _2 X^{\ell,0}_{AB})=(2r\xi^\ell _1-L\xi^\ell _S)\Omega_{AB}Y^{\ell,0}+2\xi^\ell _SY^{\ell,0}_{AB}+2\xi^\ell _V X^{\ell,0}_{AB}\\
        &\delta a_a^{\ell}Y^{\ell,0}=\partial_a({A}_{\mu}\xi^{\mu})-\xi^b_{\ell}{A}_0'e_{ba}Y^{\ell,0}\>,\>\delta a^\ell  Y^{\ell,0}_A+\delta b^\ell X^{\ell,0}_A=\partial_A({A}_{\mu}\xi^{\mu})+\frac{q_M}{r^2}(\xi^\ell _S Y_{\ell,0}^B+\xi^\ell _V X_{\ell,0}^B) e_{BA}
    \end{aligned}
    \label{eq:gauge_transf_graviton}
\end{equation}
Notice that the projection of ${A}_{\mu}\xi^{\mu}$ onto $Y_{\ell m}$ is not diagonal in $\ell$. This is entirely due to the magnetic component of the background gauge potential:
\begin{equation}
    {A}_{\mu}\xi^{\mu}={A}_a\xi_\ell^a Y_{\ell,0}+\frac{{A}_A}{r^2}(\xi^\ell _S Y^A_{\ell,0}+\xi^\ell _V X^A_{\ell,0})={A}_a\xi_\ell^a Y_{\ell,0}-\frac{q_M}{r^2}\xi^\ell _V\left(\ell\,Y_{\ell,0}+\sum_{j=1}^{\lfloor \ell/2\rfloor}\sqrt{(2\ell+1)(2(\ell-2j)+1)} Y_{\ell-2j,0}\right)
    \label{eq:gauge_transf_photon}
\end{equation}
therefore, a linear diffeomorphism with angular momentum $\ell$ generates also lower multipoles in $\delta A_{\mu}$. On the other hand, the gauge field transforms diagonally under a $U(1)$ gauge transformation
\begin{equation}
    \delta a_a^\ell =\partial_a \Lambda^\ell \>,\>\delta a^\ell =\Lambda^\ell \>,\>\delta b^\ell =0
\end{equation}
Therefore, one can combine a linear diffeomorphism with a $U(1)$ gauge transformation with parameter $\Lambda=\sum_\ell \Lambda_\ell Y^{\ell,0}$ and choose, for any fixed $\ell$, 
\begin{equation}
    \Lambda_\ell=\lambda_\ell + \frac{q_M}{r^2}\sum_{k>\ell,k-\ell\,{\rm even}}\xi^k_V\sqrt{(2\ell+1)(2k+1)}
\end{equation}
in order to get rid of the contamination generated by the linear diffeomorphism. Here the sum runs over all source multipoles $k$ present in the diffeomorphism whose lower-harmonic expansion contains $Y^{\ell,0}$. When this is done, the combined transformation of the gauge field becomes diagonal in the spherical harmonics basis
\begin{equation}
    \delta a_a^\ell =\partial_a \left({A}_b \xi^b_\ell-\ell\frac{q_M}{r^2}\xi^\ell _V+\lambda_\ell \right)-\xi^b_\ell e_{ba}{A}_0'\>,\>\delta a^\ell ={A}_b \xi^b_\ell+\lambda_\ell+(1-\ell)\frac{q_M}{r^2}\xi_V^\ell \>,\>\delta b^\ell =-\frac{q_M}{r^2}\xi_S^\ell 
\end{equation}
Before extracting the quadratic action for the above perturbations, it is convenient to pick a gauge.  For multipoles
$\ell\geq2$, a relatively simple choice is given by
\begin{equation}\label{eq:algebgauge}
w_0^\ell =0\,,\qquad K^\ell =0\,,\qquad G^\ell =0\,,\qquad h_2^\ell =0\,,\qquad a^\ell =0\,.
\end{equation}
It is easy to see that the above choice completely exhausts the gauge freedom and it is realized by picking
\begin{equation}
    \xi_V^\ell =-\frac{r^2}{2}h_2^\ell \,,\,\xi_S^\ell =-\frac{r^2}{2}G^\ell \,,\,\xi_1^\ell =-\frac{r }{2}K^\ell  +\frac{L}{2r} \xi^\ell _S\,,\,\xi_0^\ell =-\frac{w_0^\ell +\dot{\xi}^\ell _S}{f}\,,\,\lambda_\ell=-a^\ell -{A}_b \xi^b_\ell-(1-\ell)\frac{q_M}{r^2}\xi_V^\ell 
\end{equation}
or equivalently
\begin{equation}\label{gauge2}
\begin{aligned}
    & \xi_V^\ell =-\frac{r^2}{2}h_2^\ell \,,\,\xi_S^\ell =-\frac{r^2}{2}G^\ell \,,\,\xi_1^\ell =-\frac{r }{2}K^\ell  -\frac{Lr}{4}G^\ell \,,\,\xi_0^\ell =-\frac{1}{f}\left(w_0^\ell -\frac{r^2}{2}\dot{G}^\ell \right)\\
    &\lambda_\ell=-a^\ell -\frac{{A}_0}{f}\left(w_0^\ell -\frac{r^2}{2}\dot{G}^\ell \right)+(1-\ell)\frac{q_M}{2}h_2^\ell 
\end{aligned}
\end{equation}

The dipole sector, $\ell=1$, requires a separate treatment.
Indeed, the tensor harmonics $Y_{AB}^{1m}$ and $X_{AB}^{1m}$
vanish identically, so that the amplitudes $G^1$ and $h_2^1$
are absent and the algebraic gauge choice above cannot be
straightforwardly specialized to $\ell=1$. We therefore adopt a
different gauge in the dipolar sector, for which some of the gauge
parameters are fixed through radial differential equations rather
than algebraically. The complete $\ell=1$ reduction is discussed
separately in Appendix~\ref{app:l1}.

\subsection{Quadratic action }

\label{subsec:action_manipulation}
Since even- and odd-parity perturbations are not generically dynamically independent on the backgrounds considered here, we retain both sectors and derive the complete quadratic action. After imposing the algebraic gauge of Eq.~\eqref{eq:algebgauge} and integrating over the angular coordinates, the second-order action can be organized as
\begin{equation}
    \label{eq:action2total}S^{(2)}=S_{A^2}+S_{hA}+S_{h^2},
\end{equation}
where the three terms collect, respectively, the contributions quadratic in the gauge-field perturbations, the mixed gauge-field--metric terms, and the contributions quadratic in the metric perturbations where\footnote{A sum over $\ell$ is understood in front of each integral, while $e_{ab}$ is the Levi-Civita tensor of the $1+1$-dimensional manifold equipped with the metric $g_{ab}$.}
\begin{equation}
\begin{aligned}
    S_{A^2}=&\frac{1}{2}\int {\rm d}^2x\, r^2\Bigg[ -\frac{1}{2}\mathcal{L}_X\mathcal{A}_{ab}\mathcal{A}^{ab}+(e^{ab}\mathcal{A}_{ab})^2\left(\frac{1}{4}\mathcal{L}_{XX}(A_0')^2+\frac{1}{4}\mathcal{L}_{YY}\frac{q_M^2}{r^4}-\frac{1}{2}\mathcal{L}_{XY}\frac{A_0'q_M}{r^2}\right)\\
    &+2Lb(e^{ab}\mathcal{A}_{ab})\left(-\frac{1}{2}\frac{\mathcal{L}_Y}{r^2}-\frac{1}{2}(\mathcal{L}_{XX}-\mathcal{L}_{YY})\frac{A_0'q_M}{r^4}+\frac{1}{2}\mathcal{L}_{XY}\left(-\frac{(A_0')^2}{r^2}+\frac{q_M^2}{r^6}\right)\right)-\frac{L^2b^2\mathcal{L}_X}{r^4}\\
    &+4L^2b^2\left(\frac{\mathcal{L}_{XX}q_M^2}{4r^8}+\frac{\mathcal{L}_{YY}(A_0')^2}{4r^4}+\frac{\mathcal{L}_{XY}q_MA_0'}{2r^6}\right)-\frac{L\mathcal{L}_X}{r^2}\left((a_a)^2+(\partial_a b)^2\right)+\frac{2\mathcal{L}_YLe^{ab}a_a\partial_b b}{r^2}\Bigg] 
    \end{aligned}
\end{equation}

\begin{equation}
\begin{aligned}
      S_{hA}=&\frac{1}{2}\int {\rm d}^2x\, r^2\Bigg[ (e^{ab}\mathcal{A}_{ab})H\left(\frac{1}{2}\mathcal{L}_XA_0' -\frac{\mathcal{L}_{XX}}{2}(A_0')^3+\frac{Y\mathcal{L}_{YY}q_M}{2r^2}+\frac{\mathcal{L}_{XY}}{8}\left(-4YA_0'+\frac{4q_M (A_0')^2}{r^2}\right)\right)\\
    &-2\mathcal{L}_XA_0'g^{ac}e^{db}\mathcal{A}_{ab}H_{cd}-L a_a z_b g^{ab}\left(\frac{2\mathcal{L}_X q_M}{r^4}\right)+Le^{ab} a_a w_b\left(\frac{2\mathcal{L}_XA_0'}{r^2}\right)-Lg^{ab}\partial_abw_b\left(\frac{2\mathcal{L}_X q_M}{r^4}\right)\\
    &+2LbH\left(-\frac{\mathcal{L}_Xq_M}{2r^4} +\frac{\mathcal{L}_{XX}(A_0')^2q_M}{2r^4}+\frac{Y\mathcal{L}_{YY}A_0'}{2r^2}-\frac{\mathcal{L}_{XY}}{8}\left(-4\frac{Yq_M}{r^4}-\frac{4(A_0')^3}{r^2}\right)\right)\\
    &-L\partial_abz_b e^{ab}\left(\frac{2\mathcal{L}_XA_0'}{r^2}\right)\Bigg]
    \end{aligned}
\end{equation}

\begin{equation}
    \begin{aligned}
        S_{h^2}=&\frac{\MP^2}{2}\int {\rm d}^2x\bigg[-\frac{r^2}{4}\nabla_a H_{bc}(\nabla^a H^{bc}-2 \nabla^b H^{ac})-\frac{L}{4}H_{ab}H^{ab}+\frac{r^2}{4}(\nabla H)^2+\frac{L}{4}H^2-\frac{r^2}{2}\nabla_a H \nabla_bH^{ab}\\
&-r\nabla_aH \nabla_br H^{ab}-\frac{L}{4}W_{ab}W^{ab}-\frac{2L}{r}w^a\nabla_aw_b \nabla^b r+\frac{2L}{r^2}(w^a\nabla_a r)^2+LH_{ab}\nabla^a w^b+\frac{Lw^a \nabla_a H}{2}\\
&-\frac{LH\nabla_a w^a}{2}-\frac{LHw^a\nabla_ar }{r}-\frac{L}{4}Z_{ab}Z^{ab}+\frac{2L}{r^2}(z^a\nabla_a r)^2-\frac{L^2}{2r^2}z_az^a-\frac{2L}{r}z^a\nabla_a z_c \partial^c r\\
        &+\frac{r^2H^2}{8}\frac{2}{\MP^2}(\mathcal{L}-Y\mathcal{L}_Y-2X\mathcal{L}_X +\mathcal{L}_{YY}Y^2)\\
        &+\frac{r^2}{2\MP^2}(\mathcal{L}_YY-\mathcal{L}+2X\mathcal{L}_X)\left(H_{ab}H^{ab}+\frac{2L}{r^2}(w_aw^a+z_az^a)\right)+\frac{1}{2\MP^2}r^2\mathcal{L}_{XY}Y H^2 (A_0')^2\\
&-r^2\frac{\mathcal{L}_X}{2\MP^2}\left(H_{ac}H_{bd}e^{ab}e^{cd}(A_0')^2+\frac{4Lw_az_be^{ab}A_0'q_M}{r^4}\right)+r^2\frac{\mathcal{L}_{XX}}{4\MP^2}H^2(A_0')^4\bigg]
    \end{aligned}
\end{equation}
where $\mathcal{A}_{ab}\equiv \partial_a a_b-\partial_b a_a\,,\,Z_{ab}\equiv \partial_a z_b-\partial_b z_a\,,\,W_{ab}\equiv \partial_a w_b-\partial_b w_a$, $H=g^{ab}H_{ab}$ and where $L\equiv \ell(\ell +1)$.\footnote{Incidentally, this action coincides with the one obtained by \cite{Taniguchi:2025bmc}, in the limit in which their scalar field is set to a constant and upon the identification $\delta A=b\>,\>\delta A_0=a_0\>,\>\delta A_1=a_1\>,\>z_0=Q_1\>,\> z_1=W\>,\>w=h_1$.}  Although the action above is rather involved, its dynamical content can be made more transparent by eliminating the two-dimensional vector perturbations in favor of scalar variables. We first consider the odd-parity metric perturbation $z_a$. In this case, the special properties of antisymmetric tensors in two dimensions allow us to perform a direct dualization and replace $z_a$ by a scalar variable $\chi$. We subsequently treat the even-parity gauge perturbation $a_a$ and show how the remaining dynamical content can likewise be encoded in a single scalar variable $V$. 
Let us then focus on $z_a$. After introducing the rescaled field $z_a=r^2v_a$, we can isolate the subset of the action that contains linear and quadratic powers of $v_a$. Defining $V_{ab}\equiv \partial_a v_b-\partial_b v_a$, the action reads
\begin{equation}\label{ActionMetricOdd}
    \begin{aligned}
        S_{v_a}&=\frac{\MP^2}{2}\int {\rm d}^2x\Bigg[-\frac{L}{4}r^4V_{ab}V^{ab}+ L r^2v_av^a \left((\partial r)^2+r\nabla^2 r-\frac{L}{2}\right)+\frac{r^4Lv_av^a}{2}\frac{2}{\MP^2}(\mathcal{L}_YY-\mathcal{L}+2X\mathcal{L}_X)+Lv_aJ^a\Bigg]
    \end{aligned}
\end{equation}
where
\begin{equation}
J^a=-2\frac{\mathcal{L}_X}{\MP^2}a^a q_M+\frac{2\mathcal{L}_X}{\MP^2}(w_be^{ab}A_0'q_M+r^2\partial_b b e^{ab} A_0')\equiv -2\frac{\mathcal{L}_X}{\MP^2}a^a q_M+\hat{J}^a\,.
\end{equation}
The kinetic term for $v_a$ can be written in first-order form by introducing an antisymmetric auxiliary tensor $B_{ab}$
\begin{equation}
    \begin{aligned}
        S_{v_a,B}&=\frac{\MP^2}{2}\int {\rm d}^2x\,L\bigg[\frac{1}{r^4}B_{ab}B^{ab}+B_{ab}V^{ab}-  \frac{r^2}{2}M_vv_av^a +v_aJ^a\bigg]
    \end{aligned}
\end{equation}
where 
\begin{equation}
    M_v=\frac{2\mathcal{L}_Xq_M^2}{r^2\MP^2}+(L-2)
\end{equation}
Notice that the first-order action above is equivalent to the original one independently of $M_v$, since integrating out $B_{ab}$ reproduces $S_{v_a}$. The subsequent elimination of $v_a$, however, requires $M_v\neq 0$. We will show later that the no-ghost condition requires $\mathcal{L}_X>0$, hence $M_v>0$. 
Integrating out $B_{ab}$ reproduces the original action, whereas integrating out $v_a$ instead gives
\begin{equation}
    v_b=\frac{J_b-2\nabla^a B_{ab}}{r^2 M_v}\,.
\end{equation}
Since an antisymmetric tensor in two dimensions contains only one independent component, we may parametrize it as 
\begin{equation}
B_{ab}\equiv\frac{e_{ab}\chi}{2} \,.
\label{eq:chi_definition}
\end{equation}
The vector perturbation $v_a$ can therefore be traded for the scalar variable $\chi$, with
\begin{equation}
v_a=\frac{J_a+e_{ab}\nabla^b \chi}{r^2 M_v}\,.
\end{equation}
Substituting this expression back into the first-order action yields the action for the field $\chi$
\begin{equation}
    \begin{aligned}
        S_{\chi}&=\frac{\MP^2}{2}\int {\rm d}^2x\,L\bigg[-\frac{1}{2r^4}\chi^2-  \frac{1}{2}\frac{(\nabla \chi)^2}{r^2 M_v}+ \frac{J^be_{ba}\nabla^a \chi}{r^2 M_v}+  \frac{J^aJ_a}{2r^2 M_v} \bigg]
    \end{aligned}
\end{equation}
We can now apply the same logic to replace the even-parity perturbations of the gauge field with an auxiliary field $V$. The subset of $S^{(2)}$ which contains linear and quadratic powers of $a_a$ can be schematically written as
\begin{equation}\label{Actionphotoneven}
   S_{a_a}=\frac{\MP^2}{2}\int {\rm d}^2x\, r^2\bigg[ \mathcal{A}_{01}^2\mathcal{F}_A+\mathcal{A}_{01}\mathcal{C}_b+\mathcal{A}_{01}\mathcal{C}_H+M_Aa_a a^a+a_a\mathcal{J}^a\bigg]
\end{equation}
where we made use of the identities $\mathcal{A}_{ab}\mathcal{A}^{ab}=-2\mathcal{A}_{01}^2\,,\,e^{ab}\mathcal{A}_{ab}=-2\mathcal{A}_{01}\,,\,{g}^{ac}e^{db}\mathcal{A}_{ab}H_{cd}=-\mathcal{A}_{01}H$ and defined
\begin{equation}
    \begin{aligned}
       &\mathcal{F}_A\equiv \frac{\mathcal{L}_X}{\MP^2}+\frac{1}{\MP^2} \left(\mathcal{L}_{XX}(A_0')^2+\frac{\mathcal{L}_{YY}q_M^2}{r^4}-\frac{2\mathcal{L}_{XY}A_0'q_M}{r^2}\right)\>,\> M_A\equiv-\frac{L(L-2)\mathcal{L}_X}{\MP^2 r^2 M_v}\\
       &\mathcal{C}_b\equiv-4Lb\left(-\frac{\mathcal{L}_Y}{2\MP^2r^2}-\frac{1}{2\MP^2}(\mathcal{L}_{XX}-\mathcal{L}_{YY})\frac{A_0'q_M}{r^4}-\frac{\mathcal{L}_{XY}(A_0')^2}{2\MP^2r^2}+\frac{\mathcal{L}_{XY}q_M^2}{2\MP^2r^6}\right)\\
       &\mathcal{C}_H\equiv-2H\left(\frac{\mathcal{L}_X}{2\MP^2}A_0' -\frac{1}{2\MP^2}\mathcal{L}_{XX}(A_0')^3+\frac{Y\mathcal{L}_{YY}q_M}{2\MP^2r^2}+\frac{\mathcal{L}_{XY}}{2\MP^2}\left(-YA_0'+\frac{q_M (A_0')^2}{r^2}\right)\right)+\frac{2}{\MP^2}\mathcal{L}_XA_0'H\\
       &\mathcal{J}^a\equiv Le^{ab}\bigg[\frac{4\mathcal{L}_Y\partial_b b}{2\MP^2r^2}-  \frac{16\mathcal{L}_X^2(2 q_M(r^2 \partial_bb A_0'))}{8\MP^4r^4 M_v}+
\frac{w_b 4\mathcal{L}_XA_0'}{2\MP^2r^2}-  \frac{16\mathcal{L}_X^2(2 q_Mw_b A_0'q_M )}{8r^4 \MP^4M_v}- \frac{4\mathcal{L}_X q_M\nabla_b \chi}{2\MP^2r^4 M_v}\bigg]
\label{eq:FA_definition}
    \end{aligned}
\end{equation}
As before, the action \eqref{Actionphotoneven} can be equivalently written in terms of a (provisional) auxiliary field $\phi$:
\begin{equation}
   S_{a_a} =\frac{\MP^2}{2}\int {\rm d}^2x\, r^2\bigg[A\phi^2+B \phi \mathcal{A}_{01} +\mathcal{A}_{01}\mathcal{C}_b+\mathcal{A}_{01}\mathcal{C}_H+M_Aa_aa^a+a_a\mathcal{J}^a\bigg]
\end{equation}
with $-B^2/(4A)=\mathcal{F}_A$. Given that $\mathcal{A}_{01}=\dot{a}_1-a_0'$, we can integrate out both components $a_a$
\begin{equation}  a_1=\frac{B\dot{\phi}+\dot{\mathcal{C}}_b+\dot{\mathcal{C}}_H-\mathcal{J}^1}{2M_Ag^{11}}\>,\qquad\qquad 
       a_0=-\frac{(r^2 (B\phi+\mathcal{C}_b+\mathcal{C}_H))'+r^2\mathcal{J}^0}{2M_Ag^{00}r^2}
\end{equation} 
We can then define the field variable 
\begin{equation}
V \equiv r^2(B \phi+\mathcal{C}_b+\mathcal{C}_H)\,,
\label{eq:V_definition}
\end{equation}
and rewrite the action \eqref{Actionphotoneven} as
\begin{equation}
   S_V =\frac{\MP^2}{2}\int {\rm d}^2x\, r^2\bigg[-\frac{1}{4\mathcal{F}_A}\left(\frac{V}{r^2}-\mathcal{C}_b-\mathcal{C}_H\right)^2-\frac{1}{4M_Ag^{11}r^4}(\dot{V}-r^2\mathcal{J}^1)^2-\frac{1}{4M_Ag^{00}r^4}(V'+r^2\mathcal{J}^0)^2\bigg]
\end{equation}
As in the previous reduction, no new degree of freedom has been introduced: the first-order formulation is classically equivalent to the original one whenever the coefficients that have been inverted are nonvanishing. In particular, zeros of $\mathcal F_A$ or of the algebraic coefficient multiplying $a_a a^a$ correspond to degenerate limits of this reduction and have to be analyzed independently (the shape of the stability conditions in this degenerate scenario is addressed in Appendix~\ref{app:degenerateD}).
After these manipulations, the quadratic action becomes the following
       \begin{equation}
    \begin{aligned}
        S^{(2)}=&\frac{\MP^2}{2}\int {\rm d}^2x\,\bigg\{-\frac{L}{2r^4}\chi^2-  \frac{L}{2}\frac{(\nabla \chi)^2}{r^2 M_v}+ L\frac{\hat{J}^be_{ba}\nabla^a \chi}{r^2 M_v}+ L\frac{2\mathcal{L}_X^2(e^{ab}w_b A_0'q_M +r^2 e^{ab}\partial_bb A_0')^2}{\MP^4r^2 M_v} \\
        &+r^2\bigg(-\frac{1}{4\mathcal{F}_A}\left(\frac{V}{r^2}-\mathcal{C}_b-\mathcal{C}_H\right)^2-\frac{1}{4M_Afr^4}(\dot{V}-r^2\mathcal{J}^1)^2+\frac{f}{4M_Ar^4}(V'+r^2\mathcal{J}^0)^2\bigg)\\
        &+r^2\bigg(- \frac{L^2b^2\mathcal{L}_X}{\MP^2r^4}+L^2b^2\left(\frac{\mathcal{L}_{XX}q_M^2}{\MP^2r^8}+\frac{\mathcal{L}_{YY}(A_0')^2}{\MP^2r^4}+\frac{2\mathcal{L}_{XY}q_MA_0'}{\MP^2r^6}\right)-\frac{L\mathcal{L}_X(\partial_a b)^2}{\MP^2r^2}\bigg)\\
        &+r^2\bigg[LbH\left( -\frac{\mathcal{L}_Xq_M}{\MP^2r^4} +\frac{\mathcal{L}_{XX}(A_0')^2q_M}{\MP^2r^4}+\frac{Y\mathcal{L}_{YY}A_0'}{\MP^2r^2}+\frac{\mathcal{L}_{XY}}{\MP^2}\left(\frac{Yq_M}{r^4}+\frac{(A_0')^3}{r^2}\right)\right)\\
&-Lg^{ab}\partial_abw_b\frac{2\mathcal{L}_X q_M}{\MP^2r^4}\bigg]-\frac{r^2}{4}\nabla_a H_{bc}(\nabla^a H^{bc}-2 \nabla^b H^{ac})+\frac{L}{4}\left(H^2-H_{ab}H^{ab}\right)+\frac{r^2}{4}(\nabla H)^2\\
&-\frac{r^2}{2}\nabla_a H \nabla_bH^{ab}-r\nabla_aH \nabla_br H^{ab}-\frac{L}{4}W_{ab}W^{ab}-\frac{2L}{r}w^a\nabla_aw_b \nabla^b r+\frac{2L}{r^2}(w^a\nabla_a r)^2+2r^2\frac{\mathcal{L}_{XX}}{8\MP^2}H^2(A_0')^4\\
&+LH_{ab}\nabla^a w^b+\frac{Lw^a \nabla_a H}{2}-\frac{LH\nabla_a w^a}{2}-\frac{LHw^a\nabla_ar }{r}+\frac{r^2H^2}{4\MP^2}(\mathcal{L}-Y\mathcal{L}_Y-2X\mathcal{L}_X +\mathcal{L}_{YY}Y^2)\\
&+\frac{r^2}{2\MP^2}(\mathcal{L}_YY-\mathcal{L}+2X\mathcal{L}_X)\left(H_{ab}H^{ab}+\frac{2Lw_aw^a}{r^2}\right)+\frac{1}{2\MP^2}r^2\mathcal{L}_{XY}Y H^2(A_0')^2\\
&-\frac{1}{2\MP^2}r^2\mathcal{L}_XH_{ac}H_{bd}e^{ab}e^{cd}(A_0')^2\bigg\}
    \end{aligned}
\end{equation}
where we used the definition of $\hat{J}^a$. At this stage the two vector perturbations $z_a$ and $a_a$ have been eliminated in favor of the scalar variables $\chi$ and $V$. The remaining task is to disentangle the constraint variables contained in the even-parity metric sector. As expected from diffeomorphism invariance, not all the components of $H_{ab}$ and $w_a$ describe independent propagating degrees of freedom.
After integrating by parts and making use of the background equations of motion, $H_0$ appears only linearly in the quadratic action. It therefore acts as a Lagrange multiplier, whose equation of motion imposes the Hamiltonian constraint on the perturbations. It is convenient to introduce the field $\psi$ through
\begin{equation}
    H_2=\frac{L}{r}(w_1-\psi)\,.
\label{eq:H2_psi}
\end{equation}
This particular redefinition is chosen so that the Hamiltonian constraint contains $w_1$ algebraically, without radial derivatives acting on it. The constraint can consequently be solved in the form
\begin{equation}
    w_1(b,V,\psi,\psi')\>.
\end{equation}
Substitution of this solution removes $w_1$ from the action and leaves $\psi$ as the dynamical variable associated with the remaining even-parity gravitational polarization.
The field $H_1$ is nondynamical as well. Its equation of motion is algebraic in $H_1$, although its source contains time derivatives of the remaining fields, and can schematically be solved as
\begin{equation}
    H_1=H_1(\dot b,\dot V,\dot\psi,\dot\psi').
\end{equation}
Eliminating $H_1$ completes the constraint reduction.

\subsection{Kinetic and mass matrices}

\label{subsec:matrices}
Upon integration by parts, and after the manipulations described above, the quadratic action contains at most two derivatives and can be expressed entirely in terms of $\vec Q=(\chi,\psi,V,b)$. On the nondegenerate branch considered here, these four variables account for the four local propagating degrees of freedom expected in Einstein gravity coupled to a $U(1)$ gauge field: two gravitational ($\chi$ and $\psi$) and two electromagnetic ($V$ and $b$) polarizations. Their identification as even or odd variables is only kinematical; on a generic dyonic background the four modes are coupled and the propagating eigenstates are mixtures of these variables. On general grounds, the quadratic action for the remaining degrees of freedom is compactly written as 
\begin{equation}
    S^{(2)}=\int {\rm d}^2x \bigg[ \frac{1}{2}\dot{Q}_i \dot{Q}_jK_0^{ij}-\frac{1}{2}Q_i' Q_j'K_1^{ij}-\frac{1}{2}M^{ij}Q_i Q_j-C^{ij}Q_i' Q_j\bigg]\,.
\end{equation}
In particular, decomposing $C^{ij}$ into its symmetric and antisymmetric parts
\begin{equation}
    C^{ij}Q_i'Q_j=(C_S^{ij}+C_A^{ij})Q_i'Q_j=-\frac{1}{2}(C^{ij}_S)'Q_iQ_j+C_A^{ij}Q_i'Q_j
\end{equation}
The symmetric part of $C^{ij}$ can thus be absorbed into a redefinition of the mass matrix, whereas its antisymmetric part represents a genuine first-order derivative mixing between the propagating fields. The reduced quadratic action above becomes
\begin{equation}\label{eq:actionforQ}
    S^{(2)}=\int {\rm d}^2x \bigg[ \frac{1}{2}\dot{Q}_i \dot{Q}_jK_0^{ij}-\frac{1}{2}Q_i' Q_j'K_1^{ij}-\frac{1}{2}\tilde{M}^{ij}Q_i Q_j-C_A^{ij}Q_i' Q_j\bigg]
\end{equation}
where $\tilde{M}^{ij}=M^{ij}-(C^{ij}_S)'$.

A useful structural property of the system can already be anticipated
 at the level of the covariant field equations. Since the matter
 stress-energy tensor is algebraic in $F_{\mu\nu}$, its linear
 variation contains at most one derivative of the gauge-field
 perturbation. Conversely, the metric dependence of the NLED equation
 introduces at most one derivative of the metric perturbation.
 Therefore, at the principal second-derivative level, the gravitational
 and electromagnetic sectors do not mix. After removing the usual
 gauge degeneracies, the physical principal symbol factorizes
 schematically as
 \begin{equation}
 {\cal P}_{\rm phys}(k)
 =
 {\cal P}_{\rm GR}(k)\oplus{\cal P}_{\rm NLED}(k).
 \label{eq:principal_factorization}
 \end{equation}
The two gravitational polarizations consequently propagate on the
 background null cone, while the nontrivial characteristic structure
 is entirely encoded in the NLED block. We provide a covariant
 derivation of Eq.~\eqref{eq:principal_factorization} in Appendix~\ref{app:Symbol}.
 The mixing among the variables $(\chi,\psi,V,b)$ is nevertheless
 generically nonvanishing at finite wavelength and in the corresponding
 polarization eigenvectors.

The propagation properties of the perturbations follow from the principal part of the
equations of motion derived from Eq.~\cref{eq:actionforQ}. In the radial eikonal limit,
$Q_i\sim q_i\,e^{-i\omega t+ik_rr}$, with $\omega$ and $k_r$ much larger than the inverse scale
over which the background varies, terms with fewer than two derivatives are subleading (see
Appendix~\cref{app:Symbol}), and nontrivial polarizations exist whenever
\begin{equation}
\det\big[\omega^2K_0-k_r^2K_1\big]=0\,.
\end{equation}
For the system under consideration one finds
\begin{equation}
K_1=f^2K_0\,,
\end{equation}
so that, whenever $K_0$ is nondegenerate, all radial characteristics satisfy
$\omega^2=f^2k_r^2$, i.e.\ $c_r^2\equiv\omega^2/(f^2k_r^2)=1$ for a locally inertial observer.
All propagating degrees of freedom, including the two NLED polarizations, therefore share the
radial null cone of the background metric.

The angular characteristics require a slightly different eikonal limit. After decomposition in spherical harmonics, angular derivatives are encoded in the eigenvalue $L=\ell(\ell+1)$.
Consequently, terms that appear algebraic in the reduced two-dimensional action can nevertheless originate from angular derivatives in the original four-dimensional theory and become leading at large $L$. In particular, the large-$L$ contributions contained in $\tilde{M}$ encode the terms relevant for angular propagation.
To isolate the purely angular characteristics, we set the radial momentum to zero and consider the limit $L,\omega\to \infty$ keeping $\omega^2/L$ fixed.
The characteristic equation is then obtained by retaining the leading large-$L$ contribution\footnote{It is important that the eikonal limit be taken on the complete determinant, rather than independently on the different matrices, since their entries can carry different powers of $L$.} in the determinant of the matrix\footnote{Notice that the term proportional to $C_A$ does not contribute to either of the limits considered above. For radial propagation it contains only one power of the large radial momentum and is therefore subleading with respect to the terms proportional to $k_r^2$, while for purely angular propagation it vanishes upon setting $k_r=0$. A generic characteristic with simultaneously nonvanishing radial and angular momentum would instead require taking the corresponding combined eikonal limit.} $\omega^2 K_0-\tilde{M}$, namely
\begin{equation}
\det\left[\omega^2K_0-\widetilde M\right]_{\rm eik}=0\>.
\label{eq:det_eik}
\end{equation}
where ${\rm eik}$ is a shorthand for the procedure to retain the leading contribution under $L,\omega\to \infty $ with $\omega^2/L$ fixed.
Introducing an angular propagation velocity through
\begin{equation}
\omega^2=\frac{L f c^2}{r^2}\,,
\label{eq:omega_squared}
\end{equation}
the characteristic equation becomes an algebraic equation for $c(r)^2$, and its roots determine the angular characteristic velocities of the different propagating modes. A negative value of $c(r)^2$ signals an angular gradient instability, whereas real and strictly positive roots correspond to stable propagating angular characteristics with respect to the background time coordinate. When $c(r)^2<0$, the corresponding frequency becomes imaginary and, at fixed radius,
\begin{equation}
    |{\rm Im}[\omega]|
\simeq
\sqrt{|c^2|f}\,\frac{\sqrt L}{r}\,.
\end{equation}
Thus, if the theory is regarded as a classical description valid at arbitrarily short distances, the growth rate is unbounded as $L\to\infty$: perturbations of increasingly short angular wavelength grow increasingly fast, with no fastest unstable mode. This behavior reflects the loss of hyperbolicity associated with a gradient instability~\cite{Hsu:2004vr, Creminelli:2019kjy} and signals that the corresponding classical initial-value problem is not well-posed. A more formal discussion in terms of the principal symbol and characteristic surfaces is given in Appendix~\ref{app:Symbol}. In the next section, we will show that regular BH cores in this class of NLED theories necessarily develop an angular gradient instability.
\section{Linear Stability}
\label{sec:stability}
In order to discuss the constraints imposed by linear stability of the perturbations, it is convenient to simplify the notation and parametrize the first derivatives of the NLED Lagrangian in terms of two functions $\zeta(r)$ and $u(r)$, defined by
\begin{equation}
\mathcal{L}_X\equiv\frac{\zeta}{2}\,,\qquad
\mathcal{L}_Y\equiv \frac{\zeta u}{2}-\frac{q_E}{q_M}\,.
\label{eq:xi_u_variables}
\end{equation}
This parametrization is particularly convenient because it is adapted to the background electromagnetic equation. Moreover, since we focus on the genuinely dyonic branch, $q_Eq_M\neq0$, and the above parametrization is well defined. The gauge potential is therefore
\begin{equation}
A_0'=\frac{q_M}{r^2}u\>,
\label{eq:u_definition}
\end{equation}
while the background electromagnetic invariants take the simple form
\begin{equation}
X=\frac{q_M^2}{2r^4}(u^2-1)\,,\qquad
Y=-\frac{q_M^2}{r^4}u\,.
\label{eq:XY_definition}
\end{equation}
\subsection{Absence of Ghosts}
We first require the absence of ghost instabilities. In a region where $f>0$, the coordinate $t$ is timelike and the matrix $K_0$ multiplying the time derivatives in \eqref{eq:actionforQ} is the kinetic matrix of the four propagating degrees of freedom. Ghost freedom therefore requires positive definiteness of $K_0$. By Sylvester's criterion, this condition is equivalent to positivity of the four leading principal minors of $K_0$. For $\ell\geq2$, they are
\begin{equation}
\begin{aligned}
&\frac{L\MP^2}{2(L-2)r^2f}>0\,, &&
\frac{L^2\MP^2\big[(L-2)\MP^2r^2+q_M^2\zeta(1+u^2)\big]}{(L-2)\,r^4\,(L+rf'-2f)^2}>0\,,\\[2pt]
&\frac{L\MP^6\big[(L-2)\MP^2r^2+q_M^2\zeta\big]}{2\zeta(L-2)f\,r^4\,(L+rf'-2f)^2}>0\,, &&
\frac{L^2\MP^8}{4f^2r^2\,(L+rf'-2f)^2}>0\,,
\end{aligned}
\label{eq:leading_minors}
\end{equation}
where we used the $tt$ component of the Einstein equations to write the common denominator
as 
\begin{equation}
    q_Eq_Mu-\MP^2r^2(L-3f+1)-r^4\mathcal L=-\MP^2r^2(L+rf'-2f)\,.
    \label{eq:denominator_identity}
\end{equation}
 
The reduction assumes that these denominators do not vanish, i.e.\ $L+rf'-2f\neq0$. Since
curvature regularity implies $f\to1$ and $rf'\to0$ as $r\to0$, one has $L+rf'-2f\to L-2>0$
for $\ell\ge2$. This degeneracy is therefore absent in a sufficiently small neighborhood of
the regular center, which is the region relevant for our argument.
Notice that the first and fourth inequalities are automatically satisfied. The nontrivial constraints therefore come from the second and third principal minors. In particular, requiring ghost freedom for all $\ell\geq2$ selects
\begin{equation}
\zeta>0
\quad\Longrightarrow\quad
{\mathcal L}_X>0
\end{equation}
Notice that the standard Maxwell regime, $\mathcal L=X$, corresponds in our conventions to $\zeta=2$ and therefore lies on the ghost-free branch (as it should).
The same physical condition applies in regions where $f<0$, although the interpretation of the kinetic matrix changes because $r$, rather than $t$, becomes timelike. Since $K_1=f^2K_0$,
the kinetic matrix associated with evolution along the radial direction is proportional to $-K_1$. For $\zeta>0$, the leading principal minors of $K_0$ alternate in sign when $f<0$, so that $-K_1$ is positive definite. Thus the ghost-free condition remains $\mathcal{L}_X>0$ on both sides of a horizon. As a final remark, the no-ghost condition $\mathcal{L}_X>0$ is equivalent to the requirement that the null energy condition is obeyed by the background solution.

\subsection{Gradient Instabilities}
\label{subsec:gradient_inst}
Next, we turn to the possible presence of angular gradient instabilities. Contrary to the absence of ghosts, which is determined by the signature of the kinetic matrix, gradient stability is a local statement about the characteristic propagation of sufficiently short-wavelength perturbations. We therefore work in the angular eikonal limit discussed in the previous section. In particular, we are interested in implementing the limit close to the coordinate origin $r=0$.

A comment is in order concerning the use of this limit near $r=0$. The characteristic analysis is always performed locally at a fixed radius $r>0$. At such a point, a perturbation with large angular momentum has a local angular wave number of order $k_{\rm ang}\sim \sqrt{L}/r$
and hence an angular wavelength
\begin{equation}
\lambda_{\rm ang}\sim\frac{r}{\sqrt L}\,.
\label{eq:ang_wavelength}
\end{equation}
The eikonal approximation requires this wavelength to be much shorter than the characteristic scales over which the background varies. For instance, for a nonvanishing background quantity \(\Phi(r)\), a sufficient local condition is
\begin{equation}
\frac{r}{\sqrt L}\left|\frac{\Phi'}{\Phi}\right|\ll1.
\label{eq:eikonal_local_condition}
\end{equation}
For a power-law behavior $\Phi\sim r^p$, the above condition becomes
\begin{equation}
\frac{|p|}{\sqrt L}\ll1\>,
\end{equation}
and can therefore be satisfied uniformly as the origin is approached by taking $L$ sufficiently large.

For more rapidly varying nonanalytic backgrounds, the large-$L$ expansion need not be uniform in $r$. For example, a profile $\Phi\sim e^{-1/r}$ gives
\begin{equation}
\frac{r}{\sqrt L}\left|\frac{\Phi'}{\Phi}\right|
\sim\frac{1}{r\sqrt L}
\end{equation}
In this case, the $r\to0$ and $L\to \infty$ limits do not commute. This does not invalidate the local eikonal analysis: at any fixed $r>0$, the condition can still be satisfied by taking $L$ sufficiently large. In what follows we therefore first determine the angular characteristics in the large-$L$ limit at an arbitrary fixed $r$, and only afterwards study their behavior as the origin is approached. In particular, no angular propagation direction needs to be defined exactly at $r=0$, where spherical coordinates degenerate.

Using Eqs.~\eqref{eq:det_eik} and~\eqref{eq:omega_squared}, the leading large-$L$ contribution to the determinant takes the form
\begin{equation}
L^4\bigg[\frac{\MP^8f^2}{4r^{10}}(c^2-1)^2\mathcal P(c^2)+O(1/L)\bigg]=0\>.
\label{eq:Appendixsymbol}
\end{equation}
 The double factor $(c^2-1)^2$ is the angular manifestation of the
 gravitational block in Eq.~\eqref{eq:principal_factorization}: the two gravitational characteristics coincide with the metric null cone.
 The remaining polynomial $\mathcal{ P}(c^2)$ is therefore the Fresnel
 characteristic polynomial associated with the two NLED polarizations  ~\cite{Obukhov:2002xa,DeLorenci:2000yh}.

Its coefficients are most transparently written in terms of the Hessian of the NLED
Lagrangian,
\begin{equation}
\mathcal H\equiv\begin{pmatrix}\mathcal L_{XX}&\mathcal L_{XY}\\ \mathcal L_{XY}&\mathcal L_{YY}\end{pmatrix},
\label{eq:Hessian}
\end{equation}
projected onto the orthogonal vectors $e_1\equiv(1,u)$ and $e_2\equiv(u,-1)$. Defining
\begin{equation}
N\equiv\zeta r^4-2q_M^2\,e_1^T\mathcal He_1\,,\qquad
D\equiv\zeta r^4+2q_M^2\,e_2^T\mathcal He_2\,,\qquad
\Gamma\equiv q_M^2\,e_1^T\mathcal He_2\,,
\label{eq:NDGamma}
\end{equation}
the nonluminal characteristics are the roots of
\begin{equation}
\zeta r^4D\,c^4-\big(ND+\Sigma\big)\,c^2+\zeta r^4N=0\,,\qquad
\Sigma\equiv\zeta^2r^8+4\Gamma^2\,,
\label{eq:Fresnel}
\end{equation}
which coincides with the characteristic equation derived covariantly in Appendix~\cref{app:Symbol}.
For $D\neq0$, the product and the sum of the roots are $c_1^2c_2^2=N/D$ and
$c_1^2+c_2^2=(ND+\Sigma)/(\zeta r^4D)$. Since $\Sigma>0$, whenever $c_1^2c_2^2>0$ the sum has
the sign of $D$, where we used the no-ghost condition $\zeta>0$. Both roots are therefore
positive if and only if
\begin{equation}
D>0\,,\qquad N>0\,.
\label{eq:NDpos}
\end{equation}
Moreover, the discriminant
\begin{equation}
\Delta=\frac{\big(ND-\zeta^2r^8+4\Gamma^2\big)^2+16\zeta^2r^8\Gamma^2}{(\zeta r^4D)^2}\ge0
\end{equation}
is non-negative regardless of the signs of $N$ and $D$, so that the roots are always real.
The conditions~\eqref{eq:NDpos} are therefore necessary and sufficient for the absence of
angular gradient instabilities.\footnote{The degenerate case $D=0$ corresponds to $\mathcal F_A=0$, where the
$\mathcal A_{01}^2$ term in Eq.~\eqref{Actionphotoneven} vanishes and the reduction to the variable $V$ is singular. In Appendix~\ref{app:degenerateD} we integrate out the perturbations $a_a$ directly and show
that the tension between regularity, absence of ghosts and gradient stability persists also
in this case.} As we will now show, they are incompatible with regularity in a
neighborhood of the origin.

A background solution defines a curve $(X(r),Y(r))$ in the $X$-$Y$ plane. Along this trajectory, the chain rule implies
\begin{equation}\label{eq:FHT}
    F=\mathcal{H} T\,,
\end{equation}
where $T\equiv (X',Y')$ and $F=(\mathcal{L}_X',\mathcal{L}_Y')$.
Thus the background evolution fixes the action of the Hessian on the tangent vector $T$. In the basis $\{e_1,e_2\}$, the tangent vector takes the form
\begin{equation}
    T=
    \frac{q_M^2}{r^5}
    \left(2e_1+se_2\right)
\end{equation}
where we have defined $ s\equiv ru'-2u$.
Equation~\eqref{eq:FHT} therefore becomes
\begin{equation}
    F=
    \frac{q_M^2}{r^5}\mathcal{H}(2e_1+se_2)\>,
\end{equation}
with $F=(\zeta'/2,(u\zeta)'/2)$.
Contracting the above relation with the vector $2e_1-se_2$ and using the symmetry of the Hessian, we obtain
\begin{equation}
    (2e_1-se_2)^TF
    =
    \frac{q_M^2}{r^5}
    \left(
        4e_1^T\mathcal{H}e_1-s^2e_2^T\mathcal{H}e_2
    \right).
\end{equation}
Using Eq.~\eqref{eq:NDGamma}, the previous equation can be rearranged to
\begin{equation}
    4N+s^2D
    =
    (4+s^2)\zeta r^4
    -2r^5(2e_1-se_2)^TF\,.
\end{equation}
Upon substituting the explicit expression for $F$ in the $\{e_{1,2}\}$ basis, the right-hand side simplifies to
\begin{equation}\label{eq:NDZidentity}
    4N+s^2D
    =
    2r^4(2Z-rZ')
\end{equation}
where $Z\equiv \zeta(1+u^2)$.
This identity directly relates the angular stability conditions to the behavior required by regularity. Indeed, if both nonluminal angular modes are stable, Eq.~\eqref{eq:NDpos} implies
\begin{equation}
    4N+s^2D>0\,.
\end{equation}
Equation~\eqref{eq:NDZidentity} then requires
\begin{equation}
\label{eq:Zcondition}
  \left(\frac{Z}{r^2}\right)'<0
\end{equation}
Ghost freedom implies
$Z=\zeta(1+u^2)>0$.
Let us first illustrate the tension with regularity within the power-law parametrization introduced above. The background Einstein equation
\begin{equation}
    r^2f''+2(1-f)
    =
    \frac{q_M^2}{\MP^2}\frac{Z}{r^2}\,.
    \label{eq:EinsteinZ}
\end{equation}
implies, near the origin, the scaling $Z\sim Z_0r^{\beta+2}$, with $\beta\geq\alpha\geq2$. Since $Z_0>0$, angular stability requires \begin{equation}
    \beta<0\>,
\end{equation} 
in direct contradiction with the regularity condition. The argument does not, however, rely on the existence of a power-law expansion. More generally, curvature regularity implies that
\begin{equation}\label{eq:regZ}
\frac{Z}{r^2}\to 0
\qquad (\text{for}\,\,\,r\to0)\>,
\end{equation}
including for regular nonanalytic backgrounds. Equations~\eqref{eq:regZ} and \eqref{eq:Zcondition}, together with the no-ghost condition $Z>0$, are mutually incompatible in a punctured neighborhood of the origin: a positive function approaching zero at $r=0$ cannot be strictly decreasing as one moves away from the origin. Indeed, for any $0<r_1<r_2$, Eq.~\eqref{eq:Zcondition} and the mean value theorem imply $Z(r_2)/r_2^2<Z(r_1)/r_1^2$. Taking $r_1\to0^{+}$ and using $Z/r^2\to0$ then yields $Z(r_2)/r_2^2\leq0$, in contradiction with the no-ghost condition $Z>0$. 

The argument above shows that regularity is incompatible with simultaneous stability of the two angular characteristics. It is useful to sharpen this result and determine what regularity implies directly for the signs of the corresponding propagation velocities.
On the nondegenerate branch, $D\neq0$, the two photon characteristics are the roots of the characteristic equation ~\eqref{eq:Fresnel}.
As previously discussed, the associated discriminant is always non-negative on this branch, which implies that the two roots are always real. For a regular core admitting the power-law scaling $Z\simeq Z_0r^{\beta+2}$, Eq.~\eqref{eq:NDZidentity} gives $
4N+s^2D\simeq-2\beta Z_0r^{\beta+6}<0$ sufficiently close to the center. If $D>0$, this immediately implies $N<0$, and therefore $c_1^2c_2^2=N/D<0$, so exactly one characteristic is negative. If $D<0$, then $N>0$ again implies that exactly one characteristic is negative, while $N<0$ implies that both are negative. Finally, if $N=0$, one root vanishes and the other is negative. Thus, sufficiently close to a power-law regular core, at least one angular photon characteristic necessarily satisfies $c^2<0$.

More generally, without assuming a power-law expansion, suppose that both characteristics were non-negative throughout some punctured neighborhood of the origin. Then $c_1^2c_2^2=N/D\ge0$ and $c_1^2+c_2^2=(ND+\Sigma)/(\zeta r^4D)\ge0$, which imply $D>0$ and $N\ge0$. Equation~\eqref{eq:NDZidentity} would then require $(Z/r^2)'\le0$, which is incompatible with $Z/r^2>0$ and $Z/r^2\to0$ as $r\to0$. Therefore, in every sufficiently small punctured neighborhood of the regular center, there exists at least one radius at which one of the angular characteristics has $c^2<0$. Thus, arbitrarily close to a regular core, the nondegenerate branch necessarily contains at least one unstable angular photon characteristic.
It is natural, at this point, to ask whether the unstable characteristic is associated with a special direction in field space.
\subsection{Kinetic soft direction and angular characteristics}
In the eikonal limit, the kinetic matrix $K_0$ acquires a block diagonal form:
\begin{equation}
    K_0^{(\rm eik)}=\begin{bmatrix}
        \frac{\MP^2}{2r^2f}&0&0&0\\
        0&2\MP^2 f & 0&0\\
        0&0&\frac{\MP^4}{2f\zeta}&-\frac{\MP^2 \mathcal{L}_Y}{f\zeta}\\
        0&0&-\frac{\MP^2\mathcal{L}_Y}{f\zeta}&\frac{4\mathcal{L}_Y^2+\zeta^2}{2f\zeta}
    \end{bmatrix}\,.
\end{equation}
Notice that none of the components along the $\psi$ and $\chi$ directions vanishes in the $r\to0$ limit, if the background is regular. We will now show that, as we approach the core, there exists a normalized direction, in the $(V,b)$ field subspace, whose kinetic norm vanishes as $r\to 0$. It will be conventient to consider the rescaled variable $\hat{V}=M_P^2V$. Let us now focus on the $2\times2$ bottom-right block of $K_0^{\rm eik}$ (which contains the information of the photon sector) and compute its eigenvalues: \begin{equation}
\lambda_{\pm}
=
\frac{
1+4\mathcal{L}_Y^2+\zeta^2\pm\sqrt{(1+4\mathcal{L}_Y^2+\zeta^2)^2-4\zeta^2}
}{4f\zeta}\,.
\end{equation}
We can exploit regularity to infer the following scalings near the core:
\begin{equation}
0<\zeta\leq Z\>,\>|\zeta u|
\leq
\frac{Z}{2}\,.
\end{equation}
This means that, in the limit $r\to0$, both $\zeta$ and $|\zeta u|$ approach zero. In particular
$\mathcal{L}_Y\to -\frac{q_E}{q_M}$.

As a consequence, we can claim that the two eigenvalues above obey the following leading behavior as $r$ approaches zero 
\begin{equation}
\lambda_{\rm stiff}
=
\frac{1+4(q_E/q_M)^2}{2f\zeta}\,,
\qquad\qquad
\lambda_{\rm soft}
=
\frac{1}{2f(1+4(q_E/q_M)^2)}\zeta\,.
\end{equation}
Notice that, since $V$ and $b$ carry different dimensions, the individual eigenvalues depend on the relative normalization of the two fields; only their scaling with $\zeta$ and the limiting null direction derived below are independent of this choice.
It is clear that regularity forces one electromagnetic kinetic eigenvalue to vanish (the soft one) and the other one to diverge (the stiff direction). 
The corresponding limiting eigenvectors are particularly simple. Up to normalization, they are given by
\begin{equation}
\hat{\mathcal{V}}_{\rm soft}
\to
\begin{bmatrix}
2q_E/q_M\\
-\MP^2
\end{bmatrix}\>,\>
\hat{\mathcal{V}}_{\rm stiff}
\to
\begin{bmatrix}
\MP^2\\
2q_E/q_M
\end{bmatrix}\,.
\end{equation}
In particular, if we parametrize them via
\begin{equation}\label{eq:paramV}
\mathcal V(\rho)
=
\begin{bmatrix}
1\\
\rho
\end{bmatrix}\>,
\end{equation}
the soft $\rho$-parameter satisfies $\rho\to\rho_{\rm soft}\equiv-\MP^2/(2q_E/q_M)$, which identifies a specific direction in the $(V,b)$ space.\footnote{It is worth mentioning that the emergence of the soft kinetic direction does not rely on the eikonal limit. Indeed, the full kinetic matrix at finite \(\ell\) develops the same asymptotic null direction, $(0,0,2q_E/q_M,-\MP^2)$, as $r\to 0$.} 
It remains to determine how this direction is related to the two angular characteristics. Let
\begin{equation}
\mathcal{V}_i
\propto
\begin{bmatrix}
1\\
\rho_i
\end{bmatrix}\>,\>i=1,2\,,
\end{equation}
denote the eigenvectors associated with the two angular velocities $c_i^2$. It is straightforward to check that, for $\Gamma\neq0$,
\begin{equation}
\rho_i
=
\frac{\MP^2}{2}
\frac{1}{\mathcal L_Y-\frac{\zeta \Gamma}{D c_i^2-\zeta r^4}}\,.
\end{equation}
Notice that the characteristic polynomial implies the following relation
\begin{equation}\label{eq:gammaeq}
(Dc_1^2-\zeta r^4)(Dc_2^2-\zeta r^4)=-4\Gamma^2\,.
\end{equation}
Taking the absolute value, the largest of the two factors on the LHS is bounded from below by $2|\Gamma|$. Equivalently,
\begin{equation}
\min\left(\bigg|\frac{\zeta \Gamma}{D c_1^2-\zeta r^4}\bigg|,\bigg|\frac{\zeta \Gamma}{D c_2^2-\zeta r^4}\bigg|\right)
\leq\frac{\zeta}{2}\,.
\end{equation}
Thus at least one of the two characteristic eigenvectors satisfies
\begin{equation}
\bigg|\frac{\zeta \Gamma}{D c_i^2-\zeta r^4}\bigg|\leq \zeta/2\,.
\end{equation}
This is crucial to establish that the norm of that specific characteristic eigenvector, $\mathcal{V}_i$, evaluated with respect to $K_0^{\rm eik}$ vanishes in the $r\to 0$ limit. Indeed, writing the associated $\rho_i$ parameter, near the core,  schematically as
\begin{equation}
   \rho_i=\frac{\MP^2}{2}\frac{1}{\mathcal{L}_Y-O(\zeta)}
\end{equation}
it is possible to see that
\begin{equation}
    \MP^2-2\mathcal{L}_Y \rho_i \sim O(\zeta)
\end{equation}
since $\mathcal{L}_Y\to -q_E/q_M\neq0$. Thus, we can compute the norm of the characteristic eigenvector $\mathcal{V}_i$ with respect to $K_0$ near the core and claim that
\begin{equation}
    \mathcal{V}_i^T K_0^{(\rm eik)}\mathcal{V}_i=\frac{1}{2\zeta f}\left[(\MP^2-2\mathcal{L}_Y \rho_i)^2+\rho_i^2 \zeta^2 \right]= \frac{1}{\zeta}\left[O(\zeta^2)+O(\zeta^2) \right]\to0
\end{equation}
Hence at least one of the two angular characteristics has a kinetic normalization that vanishes as the regular center is approached.\footnote{For $\Gamma=0$, Eq.~\eqref{eq:gammaeq} gives $
(Dc_1^2-\zeta r^4)(Dc_2^2-\zeta r^4)=0$.
If the two factors do not vanish simultaneously, the characteristic associated with the nonvanishing factor satisfies exactly $\rho_i=\MP^2/(2\mathcal L_Y)$, and hence approaches the soft kinetic direction. If both factors vanish, the two photon characteristics coincide, $c_1^2=c_2^2=\zeta r^4/D$, and the angular propagation operator becomes proportional to the kinetic matrix. Therefore, the characteristic speed does not distinguish any direction at all. The soft direction is automatically one of the allowed characteristic directions.}
This conclusion is logically independent of the gradient-instability result derived above. Regularity and ghost freedom imply that at least one angular characteristic has $c^2<0$, and that at least one angular characteristic becomes kinetically soft. However, the regularity condition is not enough to determine how these two properties are paired. If one characteristic is stable and the other unstable, the unstable mode may be either the soft or the stiff one. If both angular characteristics are unstable, at least one of them is necessarily soft. Determining the precise correspondence between angular instability and kinetic softness on the generic nondegenerate branch requires additional information about the NLED Lagrangian or the particular background solution.
The physical interpretation of these  results depends on the regime of validity of the NLED model. If the Lagrangian is treated as an exact classical theory valid all the way to the core, the angular gradient instability extends to arbitrarily short wavelengths: for a characteristic with $c^2<0$, the exponential growth rate of the unstable linear perturbation increases with $k_{\rm ang}\sim \sqrt{L}/r$. At the same time, regularity drives one electromagnetic kinetic eigenvalue to zero as the center is approached. This signals a degeneration of the quadratic Lagrangian along a definite combination of the photon perturbations. Strictly speaking, the vanishing of the quadratic kinetic coefficient does not by itself establish strong coupling, since this requires determining the scaling of the nonlinear interactions after canonical normalization. Nevertheless, unless the corresponding higher-order coefficients vanish sufficiently rapidly, the soft kinetic normalization enhances nonlinear terms and indicates a loss of perturbative control near the regular core. Importantly, the emergence of this soft direction does not by itself invalidate the linear stability analysis. At every fixed $r>0$ on the nondegenerate branch, the kinetic matrix remains nonsingular and the angular characteristics are well defined. The gradient instability is therefore a genuine property of the quadratic equations wherever the corresponding modes lie within the perturbative regime. The possible onset of strong coupling only limits how far in momentum, or how close to the core, the quadratic description can consistently be extrapolated.
If instead NLED is interpreted as a Wilsonian EFT with a finite UV cutoff $\Lambda_{\rm UV}$, only modes satisfying
\begin{equation}
k_{\rm ang}\sim\frac{\sqrt L}{r}\lesssim \Lambda_{\rm UV}
\end{equation}
belong to the regime of validity of the effective description. The formal $L\to\infty$ divergence of the instability rate should therefore not be extrapolated beyond the cutoff. This does not, however, remove the instability whenever a negative angular characteristic is already present at momenta parametrically below $\Lambda_{\rm UV}$: such modes represent genuine unstable fluctuations within the EFT. The presence of the soft kinetic direction introduces an additional limitation. After canonical normalization, interactions involving this mode are generically enhanced by inverse powers of its vanishing kinetic coefficient, so that the relevant cutoff $\Lambda_{\rm sc}(r)$ may become smaller than $\Lambda_{\rm UV}$.\footnote{Establishing the scaling of $\Lambda_{\rm sc}(r)$ requires an analysis beyond quadratic order.} The relevant range of validity of the linear stability analysis is therefore controlled schematically by the window
\begin{equation}
\Lambda_{\rm bg}(r)\ll k_{\rm ang}\ll\Lambda_{\rm pert}(r)
\equiv
\min\{\Lambda_{\rm UV},\Lambda_{\rm sc}(r)\}\,,
\end{equation}
where $\Lambda_{\rm bg}(r)$ denotes schematically the largest local inverse length scale associated with variation of any background quantity entering the perturbation equations. Whenever the unstable characteristic exists for momenta well below $\Lambda_{\rm pert}(r)$, the gradient instability is a robust prediction of the EFT. Conversely, if the perturbative window closes before such wavelengths can be reached, the quadratic analysis ceases to be predictive and the regular-core regime becomes sensitive to additional short-distance physics. In this case, the kinetic softening can restrict the domain over which the instability can be reliably followed toward the center.

\section{Conclusions}
BH singularities are generally expected to be resolved by the UV completion of gravity. While the nature of such a completion remains unknown, this expectation has long motivated the search for regular BH solutions within classical extensions of GR. A particularly remarkable example is provided by Einstein gravity coupled to nonlinear electrodynamics (NLED), for which several nonsingular BH solutions are known to exist~\cite{Ayon-Beato:1998hmi,Ayon-Beato:1999kuh,Ayon-Beato:2000mjt,Bronnikov:2000vy,Bronnikov:2022ofk,Ansoldi:2008jw,Dymnikova:2004zc,Balart:2014cga}.

In this work, we have shown that regularity comes at a dynamical price for genuinely dyonic BHs in Einstein-NLED theories. By deriving the complete quadratic action for linear perturbations and analyzing their characteristic propagation, we have found that regularity, absence of ghosts, and angular gradient stability cannot be simultaneously maintained in a sufficiently small neighborhood of the core. In particular, any regular solution that remains on the ghost-free branch must develop an angular gradient instability as the center is approached. We have also shown that regularity drives one electromagnetic kinetic eigenvalue to zero, selecting a soft photon direction, although on the generic nondegenerate branch regularity alone does not determine whether this direction coincides with the unstable angular characteristic. Together with the corresponding results for purely electric and purely magnetic configurations~\cite{DeFelice:2024seu}, this establishes a general obstruction to constructing regular and linearly stable BHs using NLED alone within an unmodified Einstein gravitational sector.

We mention, for the sake of completeness, that some of the solutions considered here possess inner horizons and are therefore subject to the well-known mass-inflation instability~\cite{Poisson:1989zz,Poisson:1990eh,Ori:1991zz,Carballo-Rubio:2018pmi,Carballo-Rubio:2021bpr}.
Since our
argument solely relies on the regularity of the center and on the absence of ghosts, it does not
depend on the global causal structure. Hence, it applies equally to inner-extremal
configurations designed to avoid or to tame mass inflation~\cite{Bonanno:2020fgp,Carballo-Rubio:2022kad} and to
horizonless solutions sharing the same regular core.

The physical interpretation of this obstruction depends on how the NLED model is assumed to apply at short distances. If the nonlinear electromagnetic Lagrangian is treated as an exact classical theory valid at arbitrarily short wavelengths, the negative squared angular characteristic velocity leads to unstable modes whose growth rate increases without bound with angular momentum, reflecting the loss of hyperbolicity of the linearized equations. If instead NLED is interpreted as an EFT with a finite UV cutoff, modes whose local wavelength falls below the corresponding microscopic scale lie outside the regime of validity of the effective description. The formal short-wavelength divergence should then not be extrapolated beyond the cutoff. Nevertheless, the cutoff does not by itself eliminate the instability: whether unstable modes already occur within the EFT regime depends on the relation between their onset scale and the UV cutoff. The simultaneous kinetic softening may introduce an additional limitation on perturbative control near the core, although establishing an associated strong-coupling scale requires an analysis beyond quadratic order. From this perspective, our result indicates that additional short-distance physics must intervene before NLED alone can provide a controlled, regular, and stable description of the BH core.

There are several possible ways to evade the assumptions underlying this no-go result. Additional degrees of freedom can qualitatively alter this conclusion. Indeed, scalar-vector theories have been shown to admit nonsingular horizonless compact objects that are free of ghosts, gradient instabilities, and, in the electric case, strong coupling~\cite{DeFelice:2024ops,DeFelice:2025vef}. Whether analogous mechanisms can support genuinely regular BHs that remain free of ghosts, gradient instabilities, and strong-coupling pathologies throughout the spacetime is, however, still an open problem.

\section*{Acknowledgements}
LC and AL warmly thank Vania Vellucci and Antonio De Felice for their important feedback on a preliminary version of the draft.

LC acknowledges insightful discussions with Matteo Della Rocca and Andrea Sanna.

The work by LC has been partially supported by the MUR FIS2 Advanced Grant ET-NOW (CUP: B53C25001080001) and by the INFN TEONGRAV initiative.
The work of AL has been funded, in part, by the French National Research Agency (ANR) under project ANR-24-CE31-1097-01.
\appendix
\section{Principal Symbol and Characteristics}\label{app:Symbol}

In this appendix we briefly review the general definition of the principal symbol of a differential operator and how it is related to the characteristics. Suppose the evolution of a physical system is described by the set of equations\begin{equation}\label{eq:appeqmotion}
\mathcal D^i{}_j Q_j=0\>,
\end{equation}
where $D^i_j$ is a matrix-valued linear differential operator of order $n$
\begin{equation}
\mathcal {D}^i_j(x,\partial)
=
\sum_{|\alpha|\leq n}
(A_{\alpha})^i_j(x)\partial_\alpha\>,
\end{equation}
and $\alpha$ is a multi-index (the weight $|\alpha|$ equals the total number of derivatives). The \textit{symbol} of the operator $D^i_j$ is defined as the polynomial obtained by replacing each derivative with a variable in the cotangent space $\partial_{\mu}\to \xi_{\mu}$
\begin{equation}
    p^i_j(x,\xi)
=
\sum_{|\alpha|\leq n}
(A_{\alpha})^i_j(x)\xi_{\alpha}\>.
\end{equation}
The \textit{principal symbol} is obtained by retaining only the highest order terms
\begin{equation}
\mathcal{P}^i_j(x,\xi)=
\sum_{|\alpha|= n}
(A_{\alpha})^i_j(x)\xi_{\alpha}\>,
\end{equation}
and it is therefore a homogeneous polynomial of degree $n$ in $\xi$.
For a system of equations, a nonzero covector $\xi$ is \textit{characteristic} whenever the principal symbol ceases to be invertible,
\begin{equation}
{\rm \det}[\mathcal{P}](x,\xi)=0\>.
\end{equation}
The set of such covectors defines the characteristic cone in the cotangent space at $x$. 
Finally, a hypersurface $\Sigma$, locally described as $\Phi(x)={\rm const.}$ 
is said to be \textit{characteristic} when its normal $n_{\mu}=\partial_{\mu}\Phi$ is characteristic:
\begin{equation}
{\rm \det}[\mathcal{P}](x,n)=0\>.
\end{equation} The physical significance of this condition can be understood directly from the equations of motion. On a non-characteristic hypersurface, the principal symbol is invertible in the normal direction and the equations can locally be solved for the highest derivatives normal to the surface. When the determinant vanishes instead, this is no longer possible for at least one linear combination of the fields. Characteristic hypersurfaces therefore determine the local propagation cones of disturbances and the causal structure associated with the differential equations of the physical system. For a second-order system,
\begin{equation}
\mathcal{D}^i_j
=
(A^{\mu\nu})^i_j(x)
\partial_{\mu}\partial_{\nu}
+(B^{\mu})^i_j(x)\partial_{\mu}
+C^i_j(x)\>,
\end{equation}
the principal symbol reduces to
\begin{equation}
\mathcal{P}^i_j=(A^{\mu\nu})^i_j(x)\xi_{\mu}\xi_{\nu}\>,
\end{equation}
and the characteristic equation is
\begin{equation}
{\rm \det}\left[
A^{\mu\nu}(x)\xi_{\mu}\xi_{\nu}
\right]=0\>.
\end{equation}
Notice in particular that terms containing fewer than two derivatives of the perturbations do not affect the characteristic surfaces, even though they can of course affect the dynamics at finite wavelength. 
The same characteristic equation arises naturally in the short-wavelength limit. Consider the WKB ansatz
\begin{equation}
Q_i(x)=
q_i(x)
\exp\left[\frac{i}{\epsilon}\Theta(x)\right]\>,\>
\epsilon\ll1\>,
\end{equation}
where the phase varies parametrically faster than both the amplitudes $q_i(x)$ and the background coefficients entering the differential operator. Defining the local wave covector $k_{\mu}\equiv \partial_\mu\Theta$, one obtains
\begin{equation}
\partial_\mu Q_i
=
e^{i\Theta/\epsilon}
\left[
\frac{i}{\epsilon}k_\mu q_i
+\partial_\mu q_i
\right]\>,
\end{equation}
while\footnote{More generally, a term containing $m$ derivatives acting on the rapidly varying phase scales as $\epsilon^{-m}$. Thus, for an operator of order $n$, its principal part gives the leading contribution at order $\epsilon^{-n}$.}
\begin{equation}
\partial_\mu\partial_\nu Q_i
=
e^{i\Theta/\epsilon}
\left[
-\frac{1}{\epsilon^2}k_\mu k_\nu q_i
+\mathcal O(\epsilon^{-1})
\right]\>.
\end{equation}
Substituting the above profiles into the equations of motion \eqref{eq:appeqmotion} therefore gives, in the eikonal limit,
\begin{equation}
\mathcal{P}^i_j(x,k)q_j=0\>.
\end{equation}
For a nonvanishing polarization to exist, we require
\begin{equation}
\det[\mathcal{P}(x,k)]=0\>.
\end{equation}
The covector $k_{\mu}$ must therefore be characteristic.
This establishes the relation between the two descriptions. The characteristic equation is defined intrinsically through the principal symbol of the differential operator, independently of any WKB approximation. The eikonal limit provides instead a physical realization of the same construction: sufficiently short-wavelength perturbations probe only the highest-derivative part of the equations and their local dispersion relation is precisely the characteristic equation.
In a sufficiently small neighborhood of a given point, the background coefficients and the slowly varying WKB amplitudes can be regarded as approximately constant. The phase may then be approximated locally by a plane wave $\Theta(x)\simeq k_\mu x^\mu$ so that
\begin{equation}
Q_i(x)\simeq q_i e^{ik_\mu x^\mu/\epsilon}\>.
\end{equation}
For a stationary background, writing
\begin{equation}
k_\mu dx^\mu=-\omega \,dt+k_i dx^i\>,
\end{equation}
the characteristic equation becomes a local algebraic relation between $\omega$ and the spatial momenta. This is the prescription used in the main text to extract the propagation velocities. We conclude by saying that, in sufficiently symmetric scenarios, the characteristic polynomial ${\rm det}[\mathcal{P}]$ can factorize into quadratic forms. In this situation, the propagation of each eigenmode can be described in terms of an effective inverse metric $G^{\mu\nu}_{\rm eff}$, with characteristic hypersurfaces corresponding to its null cone (see for instance~\cite{Davies:2021frz,Babichev:2007dw,Bruneton:2006gf})
\begin{equation}
G_{\rm eff}^{\mu\nu}\xi_\mu\xi_\nu=0\>.
\end{equation}
Different propagating modes may in general be associated with different effective metrics and therefore with different characteristic cones.
When this effective-metric description exists, gradient instabilities admit a simple geometrical interpretation. Locally, a mode with dispersion relation
\begin{equation}
\omega^2
=
c_r^2 k_{r}^2
+
c_\Omega^2 k_{\Omega}^2
\end{equation}
is associated, up to an irrelevant conformal factor, with a principal quadratic form of the schematic form
\begin{equation}
G_{\rm eff}^{\mu\nu}\xi_\mu\xi_\nu
\sim
-\omega^2
+c_r^2k_{r}^2
+c_\Omega^2k_{\Omega}^2\,.
\end{equation}
For $c_r^2,c_{\Omega}^2>0$, the effective cone has the usual Lorentzian orientation, with the background time coordinate defining a valid local evolution direction. However, if $c_\Omega^2$ becomes negative, the effective metric changes signature, and the Cauchy problem with respect to the background time direction ceases to be hyperbolic. An angular gradient instability can therefore be viewed, whenever the effective-metric picture applies, as a signal of the loss of hyperbolicity with respect to the physical time evolution.

For the theory considered in this work, the principal symbol admits a useful simplification. The stress-energy tensor is algebraic in $F_{\mu\nu}$, so that its linear variation contains at most one derivative of the gauge perturbation, while the metric dependence of the NLED equation produces at most one derivative of the metric perturbation. Therefore, the gravitational and electromagnetic sectors decouple at the level of the principal symbol. Let us now recall the equation of motion of the electromagnetic sector Eq.~\eqref{eq:EinsteinMaxwell} and define, at principal order,\begin{equation}
 p^\nu\equiv k_\mu F^{\mu\nu}\,,\qquad
 q^\nu\equiv k_\mu\tilde F^{\mu\nu},
 \end{equation}
 and
\begin{equation}
 \mu\equiv p^\mu \delta A_\mu,
 \qquad
 \eta\equiv q^\mu \delta A_\mu\,.
 \end{equation}
 Then, using $\delta X=-\mu$, $\delta Y=-\eta$ and the linearized Bianchi identity, the electromagnetic principal equations can be reduced to
 \begin{equation}
 \left[
 {\cal L}_X k^2\,{\bf 1}
 -\mathcal{G}(k)\mathcal{H}
 \right]
 \begin{pmatrix}
 \mu\\
 \eta
\end{pmatrix}
 =0,
 \label{eq:nled_principal}
 \end{equation}
 where
 \begin{equation}
 \mathcal{H}=
 \begin{pmatrix}
 {\cal L}_{XX}&{\cal L}_{XY}\\
 {\cal L}_{XY}&{\cal L}_{YY}
 \end{pmatrix},
 \qquad
 \mathcal{G}(k)=
 \begin{pmatrix}
 p^2&p\cdot q\\
 p\cdot q&q^2
 \end{pmatrix}.
 \end{equation}
 Hence the NLED characteristic equation is
 \begin{equation}
 \det\left[
 {\cal L}_X k^2\,{\bf 1}
 -
 \mathcal{G}(k)\,\mathcal{H}
 \right]=0 .
 \label{eq:nled_fresnel}
 \end{equation}
 For purely angular propagation on a dyonic background, employing Eq.~\eqref{eq:omega_squared}, Eq.~\eqref{eq:nled_fresnel}
 reduces, up to an irrelevant nonvanishing overall factor, to
 \begin{equation}
 \zeta r^4 D\,c^4
 -
 \left(ND+\Sigma\right)c^2
 +
 \zeta r^4N
 =0 ,
 \label{eq:fresnel_ND}
 \end{equation}
where $N$, $D$ and $\Sigma$ are defined in Eqs.~\eqref{eq:NDGamma} and~\eqref{eq:Fresnel}.
This is exactly Eq.~\eqref{eq:Fresnel}, obtained in Sec.~\ref{subsec:gradient_inst} from the reduced quadratic action.

\section{Degenerate case $D=0$}
\label{app:degenerateD}

The analysis of Sec.~\ref{sec:stability} assumes that the constraint reduction leading
to the variable $V$ is nondegenerate. There is, however, a special
case which requires a separate treatment. Using Eqs.~\eqref{eq:u_definition}--\eqref{eq:XY_definition}, the
coefficient $\mathcal{F}_A$ defined in Eq.~\eqref{eq:FA_definition} can be written as
\begin{equation}
2\MP^2 r^4 \mathcal{F}_A
=
\zeta r^4
+
2q_M^2
\left(
u^2 {\cal L}_{XX}
-2u{\cal L}_{XY}
+{\cal L}_{YY}
\right)
= D ,
\label{eq:D_FA}
\end{equation}
where $D$ is precisely the quantity defined in Eq.~\eqref{eq:NDGamma}.
Therefore,
$D=0
\,\Longleftrightarrow\,
\mathcal{F}_A=0$. 
On this locus, the term quadratic in the two-dimensional field strength
$\mathcal{A}_{01}$ vanishes, and the transformation from the gauge perturbations
$a_a$ to the variable $V$ used in Sec.~\ref{sec:linear_pert} is singular. The degenerate
case must consequently be analyzed directly from the quadratic action
before $V$ is introduced.


Consider the part of the quadratic action involving the even-parity
two-dimensional gauge perturbations $a_a$. Before the introduction of
$V$, it has the schematic form given in Eq.~\eqref{Actionphotoneven},
\begin{equation}
S_{a_a}^{}
=
\frac{\MP^2}{2}\int {\rm d}^2x\, r^2\bigg[ \mathcal{A}_{01}^2\mathcal{F}_A+\mathcal{A}_{01}\mathcal{C}_b+\mathcal{A}_{01}\mathcal{C}_H+M_Aa_a a^a+a_a\mathcal{J}^a\bigg].
\label{eq:Saa_deg_start}
\end{equation}
On the degenerate branch $\mathcal{F}_A=0$, the first term is absent. Defining
$ \mathcal{C}\equiv \mathcal{C}_{b}+\mathcal{C}_H$, the remaining derivative dependence on $a_a$ is
linear,
\begin{equation}
r^2 \mathcal{C} \mathcal{A}_{01}
=
r^2 \mathcal{C}\left(\dot a_1-a_0'\right)
=
-r^2  \dot{\mathcal{C}}\,a_1
+(r^2\mathcal{C})'a_0 +\left(\text{boundary terms}\right)\,.
\label{eq:deg_ibp}
\end{equation}
Thus the two components $a_0$ and $a_1$ enter the action algebraically after an integration by parts and can be eliminated directly, without
introducing the auxiliary variable $V$.

The odd gravitational vector can still be dualized in terms of the
field $\chi$ as in Sec.~\ref{sec:linear_pert}. After eliminating the two components
$a_a$ and the odd vector constraints, the remaining metric
constraints are treated the same way as in the nondegenerate
case. In particular, $H_0$ remains a Lagrange multiplier. Introducing again the field rotation of Eq.~\eqref{eq:H2_psi},
the $H_0$ equation contains $w_1$ algebraically and can be solved for
it. The field $H_1$ is likewise nondynamical and is eliminated through
its algebraic equation of motion.

Away from further degeneracies of these algebraic constraints, the
quadratic action can therefore be reduced to three finite field
variables $Q_{\rm deg}=\left(\chi,\psi,b\right)$. 
Up to integrations by parts, we write the reduced action as
\begin{equation}
S^{(2)}_{\rm deg}=\int {\rm d}^2x\left[\frac12\dot Q_{{\rm deg},i}\dot Q_{{\rm deg},j}\,K_0^{{\rm deg}\,ij}
-\frac12Q'_{{\rm deg},i}Q'_{{\rm deg},j}\,K_1^{{\rm deg}\,ij}
-\frac12M^{{\rm deg}\,ij}Q_{{\rm deg},i}Q_{{\rm deg},j}
-C^{{\rm deg}\,ij}Q'_{{\rm deg},i}Q_{{\rm deg},j}\right]\,.
\label{eq:Sdeg_reduced}
\end{equation}
The absence of $V$ reflects the degeneracy of the quadratic gauge sector at $\mathcal{F}_A=0$. No statement about the nonlinear fate of the corresponding generic mode is needed for the analysis below; we only require regularity of the quadratic system obtained directly on the degenerate background.

Before studying the kinetic matrix, it is useful to derive two
relations obeyed by the Hessian of the NLED Lagrangian on the
degenerate branch. From $D=0$ and Eq.~\eqref{eq:NDGamma} one immediately obtains
\begin{equation}
e_2^{T}\mathcal{H} e_2
=
-\frac{\zeta r^4}{2q_M^2}\,\Longrightarrow {\cal L}_{YY}
=
2u{\cal L}_{XY}
-u^2{\cal L}_{XX}
-\frac{\zeta r^4}{2q_M^2}.
\label{eq:e2He2_deg}
\end{equation}
A second useful relation follows from the evolution of the background
in the $(X,Y)$ plane. 
Using the definitions of $T$, $F$, and $s$ from Sec.~\ref{sec:stability}, one obtains
\begin{equation}
-\frac{\zeta u'}{2}
=
\frac{q_M^2}{r^5}
\left(
2e_1^T\mathcal{H}e_2+s\,e_2^T\mathcal{H}e_2
\right)\,,
\label{eq:e2F_deg}
\end{equation}
which, together with Eq.~\eqref{eq:e2He2_deg} yields
\begin{equation}
e_1^T\mathcal{H}e_2
=
-\frac{\zeta u r^4}{2q_M^2}.
\label{eq:e1He2_deg}
\end{equation}
Expanding $e_1^T\mathcal{H}e_2$ in terms of the Hessian components and
using Eq.~\eqref{eq:e2He2_deg}, we finally obtain
\begin{equation}
{\cal L}_{XY}
=
u{\cal L}_{XX}
+
\frac{u\,\zeta r^4}
     {q_M^2(1+u^2)}\,.
\label{eq:LXY_deg}
\end{equation}
Equations~\eqref{eq:e2He2_deg} and \eqref{eq:LXY_deg} are exact
relations along the degenerate background and will considerably
simplify the kinetic and characteristic matrices.

For $f>0$, the absence of ghosts requires the kinetic matrix $K_0^{\rm deg}$ to be positive
definite. With $\chi$ normalized as in Sec.~\ref{sec:stability}, and using Eqs.~\eqref{eq:e2He2_deg}--\eqref{eq:LXY_deg} together with the same form of the common denominator as in Eq.~\eqref{eq:denominator_identity}, its three leading principal minors read
\begin{equation}
\frac{L\MP^2}{2(L-2)r^2f}>0\,,\qquad
\frac{L^2\MP^2\big[(L-2)\MP^2r^2+q_M^2Z\big]}{(L-2)\,r^4\,(L+rf'-2f)^2}>0\,,\qquad
\frac{L^3\MP^4\,Z}{2f\,r^2\,(L+rf'-2f)^2}>0\,.
\end{equation}
The first two coincide with the corresponding minors of Eq.~\eqref{eq:leading_minors}, so that the degenerate branch differs from the generic one only in the third. For $\ell\ge2$ the first minor is positive wherever $f>0$ and, as discussed below Eq.~\eqref{eq:leading_minors}, $L+rf'-2f$ does not vanish in a
sufficiently small neighborhood of the regular center. The third minor then has the sign of
$Z=\zeta(1+u^2)$, so the positive definiteness of $K_0^{\rm deg}$ requires $\zeta>0$.
Conversely, if $\zeta>0$, all three minors are positive. In the degenerate branch $D=0$, the
condition $\zeta>0$ is therefore both necessary and sufficient for the absence of ghosts.

The NLED characteristic polynomial derived in Appendix~\ref{app:Symbol} was obtained without dividing by $D$ and therefore remains valid on the degenerate locus. Setting $D=0$ in Eq.~\eqref{eq:fresnel_ND} gives
$\Sigma c^2=\zeta r^4N$. Moreover, Eq.~\eqref{eq:e1He2_deg} implies that $\Sigma =\zeta r^8 Z$. Hence, one has
\begin{equation}
    c^2_{\rm deg}=\frac{N}{r^4Z}\,.
\end{equation}
The absence of an angular gradient
instability in the remaining nonluminal mode therefore requires $N\geq0$. On the degenerate branch $D=0$, the exact identity Eq.~\eqref{eq:NDZidentity} reduces to $4N=2r^4(2Z-rZ')$, and hence
\begin{equation}
\left(\frac{Z}{r^2}\right)'\leq0 .
\end{equation}
This is incompatible with curvature regularity,
$Z/r^2\to0$ as $r\to0$, together with $Z>0$ in a punctured neighborhood of the origin. We therefore conclude that the degenerate branch does not evade the no-go result: regularity,
absence of ghosts and angular gradient stability cannot be
simultaneously satisfied. 
\section{Dipole perturbations}
\label{app:l1}

In this appendix, we discuss separately the dipole sector,
$\ell=1$. In this case the tensor harmonics on $S^2$ vanish
identically,
\begin{equation}
Y^{1m}_{AB}=X^{1m}_{AB}=0,
\end{equation}
so that the corresponding amplitudes $G$ and $h_2$ are absent.
The gauge choice adopted for $\ell\geq2$ in Eq.~\eqref{eq:algebgauge} therefore
cannot be straightforwardly specialized to the dipole sector.

A convenient gauge, which we use throughout this appendix, is
\begin{equation}
w_0^1=0,\qquad
H_2^1=0,\qquad
K^1=0,\qquad
z_1^1=0,\qquad
a^1=0.
\label{eq:dipolar_gauge}
\end{equation}
Unlike the gauge employed for $\ell\geq2$, the conditions
$H_2^1=0$ and $z_1^1=0$ are implemented through differential
gauge transformations. In particular, using Eqs.~\eqref{eq:gauge_transf_graviton} and~\eqref{eq:gauge_transf_photon},
the polar gauge parameters satisfy
\begin{equation}
\xi_1^1 =
-\frac{r}{2}K^1+\frac{\xi_S^1}{r}\,,\qquad\qquad
\xi_0^1 =
-\frac{w_0^1+\dot{\xi}_S^1}{f}\,,\qquad\qquad
H_2^1+2(\xi_1^1)'
-\frac{f'}{f}\xi_1^1 =0 \,,
\end{equation}
where the last equation determines $\xi_S^1$ through a
first-order radial differential equation. Equivalently,

\begin{equation}
(\xi_S^1)'
-\left(\frac{1}{r}+\frac{f'}{2f}\right)\xi_S^1
=
\frac{r^2}{2}(K^1)'
+\left(\frac{r}{2}-\frac{r^2f'}{4f}\right)K^1
-\frac{r}{2}H_2^1 .
\end{equation}

The axial diffeomorphism and the $U(1)$ gauge transformation are
instead fixed by

\begin{equation}
\left(\frac{\xi_V^1}{r^2}\right)'
=
-\frac{z_1^1}{r^2},
\qquad
\lambda^1
=
-\left(a^1+A_b\xi^{1,b}\right).
\end{equation}

The differential nature of the dipolar gauge fixing is important
when deriving the reduced equations. In practice, $H_2^1$ and
$z_1^1$ are retained when varying the quadratic action, and the
conditions $H_2^1=z_1^1=0$ are imposed only afterwards.

After deriving the equations of motion and subsequently imposing the
gauge conditions, the equations associated with
$a_0^1$, $a_1^1$, $H_0^1$ and $H_1^1$ form a system of
nondynamical constraints. Away from the degenerate loci discussed in
Appendix~\ref{app:degenerateD}, this system can be solved algebraically for
$a_0^1$, $a_1^1$, $w_1^1$ and $H_1^1$, in terms of $\chi,V,b^1$ and their first derivatives.
The explicit expressions are lengthy and will not be reported here.
The equations originating from the two components of the odd
gravitational vector contain further information. In particular, the
equation associated with $z_1^1$ is retained even though
$z_1^1=0$ in our gauge, since the variation is performed before the
gauge condition is imposed. After substituting the solutions of the equation of motion for $a_0^1$, $a_1^1$, $w_1^1$ and $H_1^1$, the
$z_0^1$ and $z_1^1$ equations reduce to two first-order equations for
the same combination of the remaining fields. We employ again the definitions of the field variables $\chi^1$ and $V^1$, introduced in Eqs.~\eqref{eq:chi_definition} and~\eqref{eq:V_definition}.
Then, defining
\begin{equation}
 {\cal K}
 \equiv
 \chi^1+\frac{q_M}{2}V^1
 +\frac{2r^2A_0'{\cal L}_X}{\MP^2}b^1\,, 
\end{equation}
the two residual equations take the simple form
\begin{equation}
 \dot{\cal K}=0,
 \qquad\qquad
 {\cal K}'=0 .
\end{equation}
They can therefore be integrated to give
\begin{equation}
 \chi^1+\frac{q_M}{2}V^1
 +\frac{2r^2A_0'{\cal L}_X}{\MP^2}\,b^1
 ={\cal C}\,,
 \label{eq:constraint_chi}
\end{equation}
where ${\cal C}$ is an integration constant.
Equation~\eqref{eq:constraint_chi} shows that $\chi^1$ does not provide an independent local dipolar degree of freedom. It can be eliminated in favor of $V^1$ and $b^1$.
The integration constant ${\cal C}$ does not contribute to the kinetic matrix and is therefore irrelevant for the local no-ghost analysis. 
After eliminating $\chi^1$, the dipolar quadratic action contains only
two independent local propagating variables, which we collect as $\vec{Q}_{\rm dip}=(V^1,b^1)$. These two remaining local modes constitute the two vector degrees of freedom of the theory; on a dyonic background their perturbative representatives are generically mixtures of metric and electromagnetic variables.

Up to integrations by parts, the reduced action can be written as
\begin{equation}
 S^{(2)}_{\ell=1}=\int {\rm d}^2x\left[\frac12\dot Q_{{\rm dip},i}\dot Q_{{\rm dip},j}\,K_0^{(\ell=1)\,ij}
-\frac12Q'_{{\rm dip},i}Q'_{{\rm dip},j}\,K_1^{(\ell=1)\,ij}
-\frac12M^{(\ell=1)\,ij}Q_{{\rm dip},i}Q_{{\rm dip},j}
-C^{(\ell=1)\,ij}Q'_{{\rm dip},i}Q_{{\rm dip},j}\right].
\end{equation}
Finally, imposing that the kinetic matrix $K_0^{(\ell=1)}$ be positive definite, i.e.\ that
$\det K_0^{(\ell=1)}>0$ and $K_0^{(\ell=1)\,11}>0$ hold simultaneously, one finds that the only required condition is simply $\zeta>0$. Hence, no extra bound is imposed on the variables $(\zeta,u)$ introduced in Eq.~\eqref{eq:xi_u_variables}, apart from the one already obtained from the no-ghost condition for $\ell \geq2$. This result is consistent with the analysis presented in~\cite{Taniguchi:2025bmc}, despite the different number of degrees of freedom and gauge choice.

\bibliographystyle{apsrev4-2}
\bibliography{bibliography}
\end{document}